\documentclass[%
 aip,
 pof,
 amsmath,amssymb,
 reprint,%
]{revtex4-1}

\usepackage{graphicx}% Include figure files
\usepackage{dcolumn}% Align table columns on decimal point
\usepackage{bm}% bold math
\usepackage[utf8]{inputenc}
\usepackage[T1]{fontenc}
\usepackage{siunitx}
\usepackage{tikz}
\usetikzlibrary{plotmarks}
\usepackage{subcaption}
\usepackage{todonotes}
\usepackage{xcolor}

\newcommand*\emptytriangle[1][1.75ex]{\tikz\draw[thick,scale=0.25] (0,0) -- (0.5,0.86) -- (1,0) -- (0,0) (#1,#1);} 
\newcommand*\emptytriangledown[1][1.75ex]{\tikz\draw[thick,scale=0.25] (0,0) -- (0.5,-0.86) -- (1,0) -- (0,0) (#1,#1);} 
\newcommand*\emptysquare[1][1.75ex]{\tikz\draw[thick] (0,0) rectangle (#1,#1);}
\newcommand*\emptydiamond[1][1.75ex]{\tikz\draw[thick,rotate=45,scale=0.9] (0,0) rectangle (#1,#1);} 
\newcommand*\emptycirc[1][1ex]{\tikz\draw[thick] (0,0) circle (#1);} 

\newcommand*\filledtriangle[1][1.75ex]{\tikz\draw[thick,scale=0.25,fill=black] (0,0) -- (0.5,0.86) -- (1,0) -- (0,0) (#1,#1);} 
\newcommand*\filledsquare[1][1.75ex]{\tikz\fill[thick] (0,0) rectangle (#1,#1);}

\begin{document}

\title[Pattern formation of spherical particles in an oscillating flow]{Pattern formation of spherical particles in an oscillating flow}
% Force line breaks with \\
\author{T.J.J.M. van Overveld}
\affiliation{Fluids and Flows group and J.M. Burgers Center for Fluid Mechanics, Department of Applied Physics, Eindhoven University of Technology, P.O. Box 513, 5600 MB Eindhoven, The Netherlands}
\author{H.J.H. Clercx}
\affiliation{Fluids and Flows group and J.M. Burgers Center for Fluid Mechanics, Department of Applied Physics, Eindhoven University of Technology, P.O. Box 513, 5600 MB Eindhoven, The Netherlands}
\author{M. Duran-Matute}
\email{m.duran.matute@tue.nl}
\affiliation{Fluids and Flows group and J.M. Burgers Center for Fluid Mechanics, Department of Applied Physics, Eindhoven University of Technology, P.O. Box 513, 5600 MB Eindhoven, The Netherlands}

\date{\today}% It is always \today, today,

\begin{abstract}
We study the self-organization of spherical particles in an oscillating flow through experiments inside an oscillating box.
The interactions between the particles and the time-averaged (steady streaming) flow lead to the formation of either one-particle-thick chains or multiple-particle-wide bands, depending on the oscillatory conditions. Both the chains and the bands are oriented perpendicular to the direction of oscillation with a regular spacing between them. For all our experiments, this spacing is only a function of the relative particle-fluid excursion length normalized by the particle diameter, $A_r/D$, implying that it is an intrinsic quantity that is established only by the hydrodynamics. In contrast, the width of the bands depends on both $A_r/D$ and the confinement, characterized by the particle coverage fraction $\phi$. Using the relation for the chain spacing, we accurately predict the transition from one-particle-thick chains to wider bands as a function of $\phi$ and $A_r/D$. 
Our experimental results are complemented with numerical simulations in which the flow around the particles is fully resolved. These simulations show that the regular chain spacing arises from the balance between long-range attractive and short-range repulsive hydrodynamic interactions, caused by the vortices in the steady streaming flow. 
We further show that these vortices induce an additional attractive interaction at very short range when $A_r/D\gtrsim0.7$, which stabilizes the multiple-particle-wide bands.
Finally, we give a comprehensive overview of the parameter space where we illustrate the different regions using our experimental data.
\end{abstract}

\maketitle

\section{\label{sec:introduction}Introduction}
Granular systems are widely present in nature and often exhibit self-organization into patterns. The understanding of these patterns is essential for many industrial processes \citep{jaeger1996granular,sanchez2004stripe}.
When the granular material is immersed in a fluid, an even richer range of pattern-forming behavior is observed \citep{aranson2006patterns}. This is due to the hydrodynamic interactions that can induce (additional) non-linearities to the particle dynamics \citep{fortes1987nonlinear}, potentially leading to macroscopic effects \citep{voth2002ordered,thomas2004structures}.
Due to their rich phenomenology, the fluid-immersed patterns are relevant in maritime settings\citep{blondeaux1990sand} and systems that contain either colloids\citep{xu2016review,lotito2020pattern} or active matter\citep{zhang2022guiding}.

One specific phenomenon is the self-organization of spherical particles in an oscillating flow. In laboratory experiments, \citet{wunenburger2002periodic} found that spherical particles, submerged in a viscous fluid and subjected to horizontal oscillations, form one-particle-thick chains oriented perpendicularly to the direction of oscillation. They attributed the formation of these chains to a non-zero residual flow, known as `steady streaming', that remains after averaging over a full oscillation period\citep{riley1966sphere}. 
Additionally, they found that the particle chains form a periodic pattern with a regular spacing, which was attributed to an equilibrium between attractive forces between the chains at large distances and repulsive forces at small distances. They described the spacing using an empirical function of the particle diameter, its excursion length relative to the fluid, and the particle Reynolds number. 
However, the sub-millimeter-sized particles could not be fully resolved in the experiments, such that the underlying physical mechanisms that govern the pattern at the particle level were not completely understood. As a result, the origin of the attractive forces between the particle chains and the empirical scaling could not be determined. 

To gain a better understanding of the physical mechanisms underlying the formation of one-particle-thick chains, \citet{klotsa2007interaction} studied a pair of particles (i.e. the shortest possible chain) in an oscillating box. They found that the alignment of the two particles perpendicular to the direction of oscillation is due to steady streaming. They further identified a gap between the particles, which they described as a function of the fluid viscosity and the oscillation parameters.
Subsequently, \citet{overveld2022numerical} extended this study using numerical simulations, covering a more extensive region of the parameter space. They found that when the particle-bottom friction is negligible, the gap width solely depends on two dimensionless parameters: the relative particle-fluid excursion length and the typical viscous length scale (the Stokes boundary layer thickness), both normalized by the particle diameter.

The role of the steady streaming flow in the chain formation process was further investigated by \citet{klotsa2009chain}, who described the average flow as a set of vortex rings, with two `inner' vortices located close to each particle and two `outer' vortices surrounding the entire configuration. The interaction between the inner and outer vortices leads to the formation of stagnation points next to the particles. Other particles tend to roughly follow the streamlines of the steady streaming flow, which lead them toward one of these stagnation points.
Additionally, \citet{klotsa2009chain} demonstrated that the structure of the steady streaming flow and the associated ordering mechanism is similar for isolated particles, pairs of particles, and short chains of particles. 
However, their work did not address the characteristics of the patterns once they are formed, such as the spacing between the chains as a function of the governing parameters. Furthermore, the long-range attractive force hypothesized by \citet{wunenburger2002periodic} was not observed in the numerical simulations of \citet{klotsa2009chain}, leaving the underlying physical mechanism that governs the separation unclear.

In this study, we provide a detailed characterization of the particle chains in an oscillating flow and describe the underlying physical mechanism. We designed an experimental set-up that is significantly larger than those used in previous studies, allowing for a detailed examination of the patterns at the scales of the particles. 
Through experiments covering a broad region of the parameter space, we reveal in detail how the spacing between the chains varies as a function of the governing parameters. 
Furthermore, our experiments show that, in addition to chains, the particles can self-organize into bands that are multiple particles wide. We quantify the width of these bands and show the effect of confinement on their formation and characteristics. 
In fact, the role of confinement in steering self-organization is currently a relevant topic in many scientific disciplines \citep{araujo2023steering}.

Next, we use direct numerical simulations to fully resolve the flow around the particles and get a detailed understanding of the time-averaged flow. 
We use a numerical code\citep{breugem2012second} that is previously validated and used to study the self-organization and dynamics of particle pairs\citep{overveld2022numerical,overveld2022effect}. 
Through our simulations, we identify the physical mechanisms that cause the long-range attractive and short-range repulsive forces between chains. We show how the flow conditions affect these interactions and explain that the same physical mechanisms are responsible for the formation of the wide bands of particles as observed in our experiments. 

The study is organized as follows. In Sec.~\ref{sec:experimentalsetup}, we describe the experimental setup and measurement approach. Section~\ref{sec:results} presents our experimental results on the patterns and Sec.~\ref{sec:mechanism} illustrates the physical mechanism that drives the attractive and repulsive interactions between chains, using numerical simulations. In Sec.~\ref{sec:phasespace}, we give a comprehensive overview of the parameter space. Finally, we give our conclusions in Sec.~\ref{sec:conclusions}.

\section{\label{sec:experimentalsetup}Experimental method}
\subsection{Experimental setup}
\begin{figure*}[ht]
    \includegraphics[width=\linewidth]{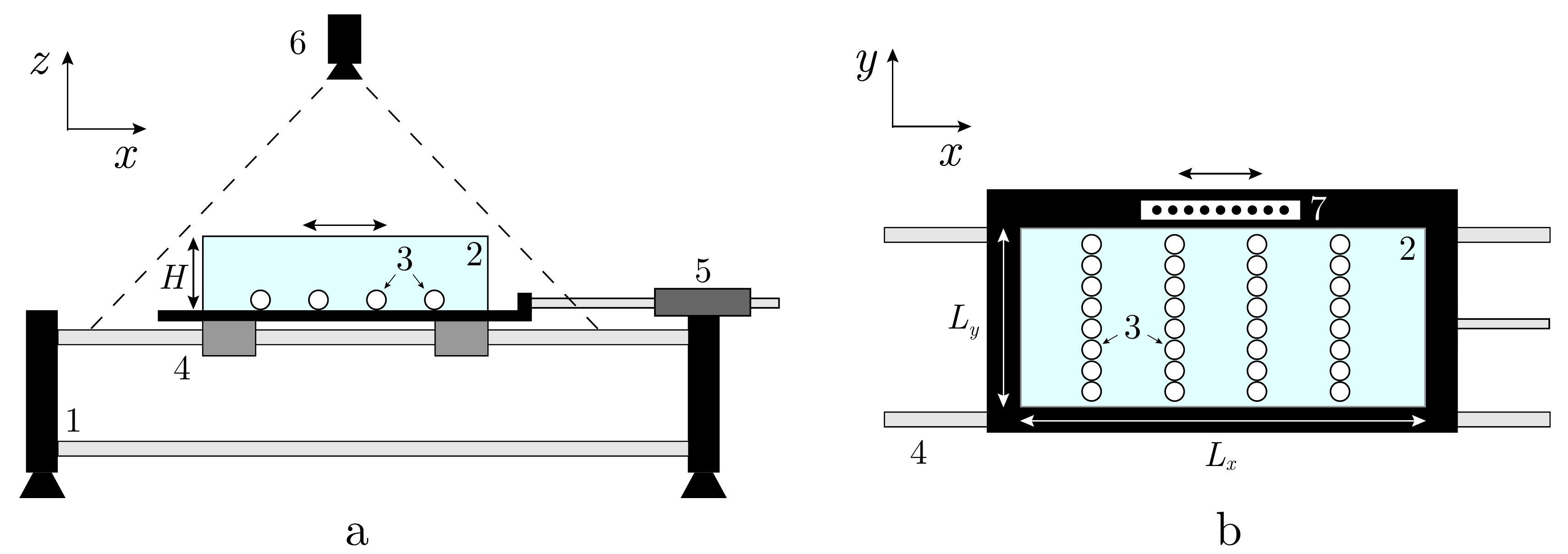}
    \caption{Schematic (a) side and (b) camera views of the experimental setup. The numbers indicate the mounting frame (1), perspex tank filled with water (2), stainless steel particles (3), guide rails (4), PID-controlled linear motor (5), camera (6), and position reference points (7). The double-headed arrows indicate the direction of oscillation, along the $x$-axis.}
    \label{fig:experimentalsetup}
\end{figure*}

A schematic representation of the experimental setup is shown in Fig.~\ref{fig:experimentalsetup}. For all experiments, we use a transparent perspex box with inner dimensions $L_x\times L_y\times H = 500 \times 250 \times \SI{55}{mm}$. The box is placed on a platform mounted on linear bearings and guide rails, which are in turn fixed to an optical table. A \SI{5.0}{mm} thick glass plate is placed inside the box, to ensure a smooth and flat bottom surface. The effective height of the box is thus \SI{50}{mm}. The plate is leveled with a precision of $0.02^\circ$ by adjusting the height of the support structure with \SI{0.1}{mm} accuracy.

The box is filled with tap water with kinematic viscosity $\nu=\left(1.05\pm0.05\right)\times10^{-6}\,\mathrm{m^2/s}$ and density $\rho_f=\left(0.999\pm0.001\right)\times10^3\,\mathrm{kg/m^3}$, then closed with a lid to eliminate any water-air interface. 
The particles are nonmagnetic, spherical, stainless steel ball bearings with density $\rho_s=\left(7.69\pm0.01\right)\times10^3\,\mathrm{kg/m^3}$. Two different sizes of particles, with diameters $D=5$ and $\SI{10}{mm}$, were used. The particles are of grade 40, which means that their diameter, roundness, and smoothness are accurate to about $\SI{1}{\micro m}$.

A PID-controlled linear motor (LinMot P01-37x120F/100x180-HP) moves the box and platform along the rails, following a user-defined sinusoidal curve that is discretized into \SI{1}{ms} increments. During oscillatory motion, the difference between the user-defined curve and the position of the box is typically less than \SI{1.0}{mm}. 
As the motor causes the box to oscillate, it creates a reaction force that leads to shaking of the optical table on which the entire setup is mounted. The amplitude of this secondary motion is typically a few millimeters but can increase up to one centimeter near the resonance frequency of the table, around $\SI{4.3}{Hz}$.
The net motion of the box in the lab frame is again a sinusoidal oscillation, which we measure by tracking reference points (with a diameter of \SI{10}{mm} and spaced \SI{30}{mm} apart) on the platform next to the box, see (7) in Fig.~\ref{fig:experimentalsetup}(b). 
These measured positions are the basis for the position of the box in the lab frame.

We define a coordinate system with the $x$-axis parallel to the direction of oscillation, the $y$-axis in the other horizontal direction, and the $z$-axis vertically up, as shown in Fig.~\ref{fig:experimentalsetup}. In the stationary lab frame, the $x$-position of the box is given by
\begin{equation}\label{eq:boxmotion}
    x_\mathrm{box} = A\sin{\left(2\pi f t+\phi_0\right)}+x_0,
\end{equation}
where $A$ is the excursion length in the lab frame, $f$ is the frequency, $t$ is the time, $\phi_0$ is an arbitrary phase of the box, and $x_0$ is an offset from a reference position.

The experiments are recorded using a RedLake MegaPlus II camera with a SONY \SI{16}{mm} f/1.8 lens, which is positioned approximately \SI{1}{m} above the box. It is worth noting that the camera is not connected to the optical table and remains stationary with respect to the lab frame. The typical resolution of the recordings is 2.2 pixels per millimeter.
The camera is triggered at an adjustable frame rate $f_\mathrm{cam}$, which enables two types of recordings. The first is a stroboscopic recording, with $f_\mathrm{cam}=f$, which is used to visualize the evolution of the patterns over many periods. The other type is taken at a higher frame rate, with $f_\mathrm{cam}$ not an integer multiple of $f$. Such a recording captures the particles at different phases of the oscillatory motion, which allows for the reconstruction of the streamwise particle motion \citep{wunenburger2002periodic,dangles2008relative}. 
In our experiments with the second type of recording, we set $f_\mathrm{cam}=\SI{20.55}{Hz}$ because it is not an integer multiple of any of the $f$ values used in our experiments.

\subsection{Measurement approach}
In our experiments, the parameter space is explored by varying $A$, $f$, $D$, and the number of particles inside the box $N$. The experiments have been divided into five series as presented in Tab.~\ref{tab:measurementseries}. For each series, $D$ and $N$ are kept constant, while the oscillatory conditions in terms of $A$ and $f$ are varied. 
The value of $N$ is chosen such that the particle coverage fraction 
\begin{equation}\label{eq:coveragefraction}
    \phi = \frac{\pi N D^2}{4 L_x L_y}
\end{equation}
is either 1, 1.5, or 2 times the lowest value considered ($\phi=0.173$).

Our primary focus in these experiments is the self-organization of the particles due to the steady streaming flows. We are not interested in the effects of variations in friction of the particles with the bottom. Therefore, we keep constant the relative strength of the particle-bottom friction compared to the driving force. Under the assumption that lift forces are negligible\citep{overveld2022numerical,overveld2022effect}, the ratio between these forces is proportional to $(s-1)\mu_c/\Gamma$, where $s=\rho_s/\rho_f$, $\mu_c$ is the Coulomb friction coefficient ($\mu_c\approx0.3$ for the interface between glass and stainless-steel \citep{engineering_toolbox}), and
\begin{equation}
    \Gamma=\frac{A\left(2\pi f\right)^2}{g}
\end{equation}
is the ratio between the maximum oscillatory acceleration and the gravitational acceleration $g=\SI{9.81}{m^2/s}$. In our experiments, $\Gamma\approx0.75$ to keep the relative importance of particle-bottom friction similar for cases with different flow conditions.
As a result of fixing $\Gamma$, we must simultaneously vary the amplitude and frequency. We use frequencies in the range $2.0-\SI{5.7}{Hz}$, which correspond to amplitudes between $5.74-\SI{46.6}{mm}$.
However, due to the shaking of the optical table, the values of $A$ in the lab frame are usually slightly larger than the user-defined motor amplitudes, such that the values of $\Gamma$ range from $0.75$ to $0.93$. 
We show that this increase in $\Gamma$ does not have a significant impact on the particle motion in Sec.~\ref{sec:streamwise} and Appendix~\ref{sec:appendix}. 

Based on the control parameters, we define two dimensionless quantities that are important for the particle dynamics in oscillatory flows are the normalized excursion length $A/D$ and the normalized viscous length scale 
\begin{equation}
    \frac{\delta}{D}=\frac{\sqrt{\nu/\left(\pi f\right)}}{D},
\end{equation}
where $\delta$ is commonly known as the Stokes boundary layer thickness. For a pair of particles in an oscillating flow, $\delta/D$ determines the gap width between the particles in the case of small particle-fluid excursion lengths \citep{overveld2022numerical}.
A consequence of fixing $\Gamma$ while varying both $A$ and $f$ is that also $A/D$ and $\delta/D$ are co-varied ($Af^2=\mathrm{Constant}\implies (\delta/D)\propto (A/D)^{1/4}$).

\begin{table*}[ht]
    \caption{Overview of the parameters that are varied between experiment series. Each series includes multiple experiments in which the frequency and the amplitude of the oscillations are varied.}
    \label{tab:measurementseries}
    \begin{tabular}{c|c|c|c|c|c|c}
        \shortstack{Experiment \\ series} & Symbol & \shortstack{Particle diameter \\ $D$ [mm]} & \shortstack{Number of \\ particles $N$} & \shortstack{Particle coverage \\ fraction $\phi$} & \shortstack{User-defined \\ frequency $f$ [Hz]} & \shortstack{Measured amplitude \\in lab frame $A$ [mm]} \\ \hline
        A1 & \emptytriangle & 10 & 275  & 0.173 & 2.0-5.6 & 6.15-47.9 \\
        A2 & \emptytriangledown & 10 & 413  & 0.259 & 2.0-5.5 & 6.51-48.2 \\
        A3 & \emptysquare & 10 & 550  & 0.346 & 2.0-5.6 & 6.15-48.0 \\
        A4 & \emptydiamond & 10 & 688  & 0.432 & 2.0-5.2 & 7.14-48.3 \\
        A5 & \emptycirc & 10 & 825  & 0.518 & 2.0-5.1 & 7.41-48.4 \\
        B1 & \filledtriangle & 5  & 1100 & 0.173 & 2.0-5.7 & 6.08-48.0 \\
        B2 & \filledsquare & 5  & 2200 & 0.346 & 2.0-5.7 & 5.96-47.9 \\
    \end{tabular}
\end{table*}

The positions of the particles are identified using TrackPy\cite{allan2019trackpy}, an open-source software package based on the core feature-finding and linking algorithms by \citet{crocker1996methods}.
Before each series of experiments, a calibration routine is performed using a \SI{2.5}{mm} thick plate with \SI{2.5}{mm} diameter dots, spaced \SI{5}{mm} apart in a $9\times19$ grid. The plate is positioned at the bottom of the tank, such that the calibration is performed at the mid-height of the \SI{5}{mm} particles. The mid-height of the \SI{10}{mm} particles lies \SI{2.5}{mm} above the calibration height, which leads to errors no larger than \SI{0.5}{mm}.
The recordings for the calibration are taken at different known positions of the box, corresponding to typical positions during oscillatory motion. The pixel coordinates of the dots $(x_\mathrm{pixel},y_\mathrm{pixel})$ are obtained using TrackPy. The positions of the dots in the lab frame ($x_\mathrm{real},y_\mathrm{real}$) are also known.
Then, the coefficients $P_{\alpha\beta}$ and $Q_{\alpha\beta}$ of the third-order polynomials
\begin{eqnarray}\label{eq:calibrationpolynomialx}
    x_\mathrm{real} &=& \sum_{\alpha=0}^{3}\sum_{\beta=0}^{3}P_{\alpha\beta}x_\mathrm{pixel}^\alpha y_\mathrm{pixel}^\beta, \\
    \label{eq:calibrationpolynomialy}
    y_\mathrm{real} &=& \sum_{\alpha=0}^{3}\sum_{\beta=0}^{3}Q_{\alpha\beta}x_\mathrm{pixel}^\alpha y_\mathrm{pixel}^\beta
\end{eqnarray}
are determined using a least-squares fitting routine. Note that the indices $\alpha$ and $\beta$ are here also used as exponents. This conversion also corrects for image distortion and for light refraction at the water-perspex and perspex-air interfaces. 
In the experiments with particles, we combine the pixel coordinates from TrackPy with the coefficients of Eqs.~\eqref{eq:calibrationpolynomialx}~and~\eqref{eq:calibrationpolynomialy} to determine the particle positions in the lab frame. The error in these positions is smaller than \SI{1}{mm}, typically around \SI{0.2}{mm}. This error is significantly smaller than the variations in the quantities used to describe the patterns, such as the relative particle-fluid excursion length (see Sec.~\ref{sec:streamwise}) and the spacing between chains (see Sec.~\ref{sec:distance}). The variations in these quantities of interest are typically on the centimeter scale. 

\section{\label{sec:results}Pattern formation and characteristics}
\subsection{Formation of the patterns}\label{sec:formation}
\begin{figure}[ht]
    \includegraphics[width=\linewidth]{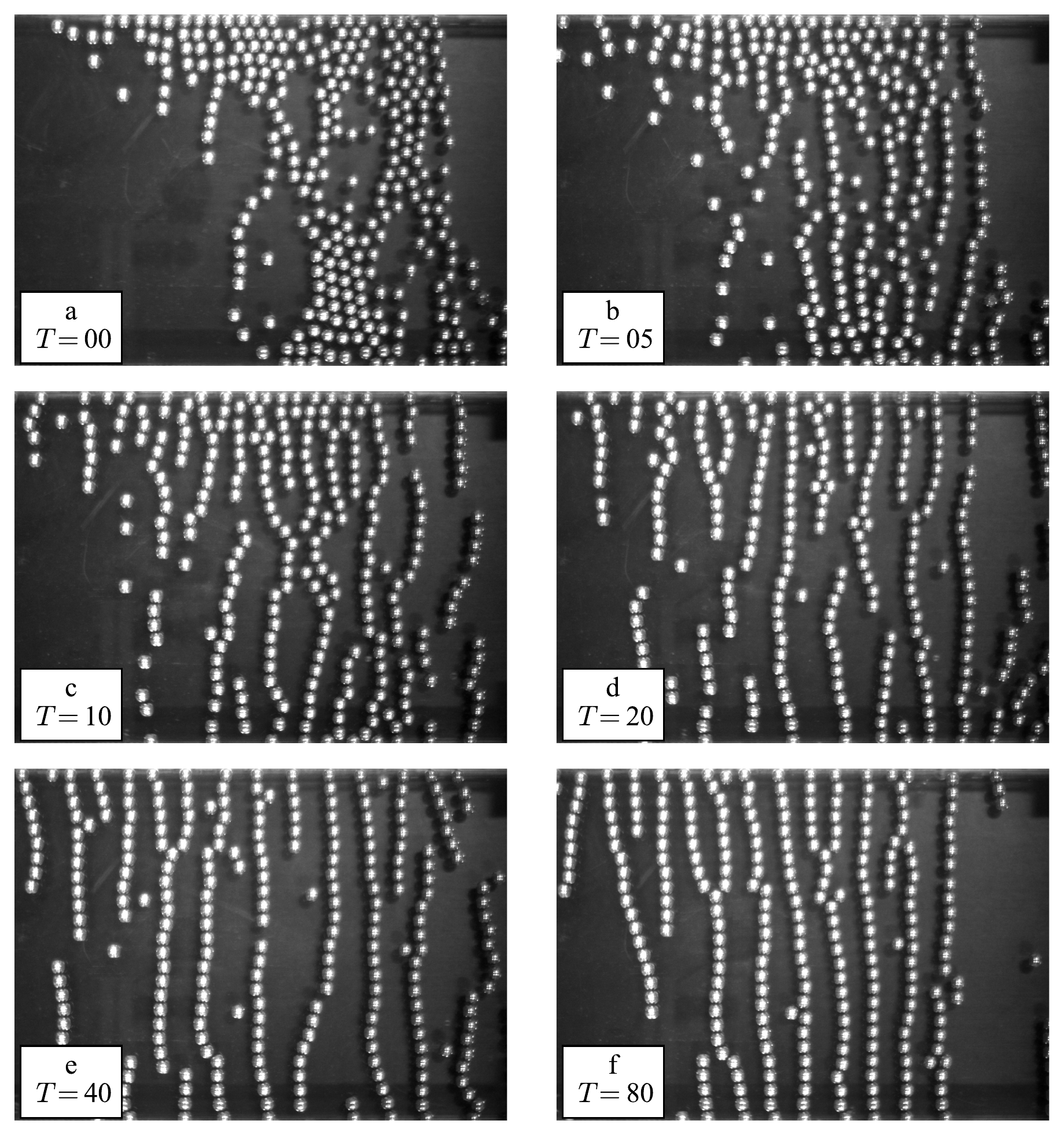}
    \caption{These images show the formation of a pattern in a part of the oscillating box for an experiment in series A1 ($f=\SI{5.0}{Hz}$, $A\approx \SI{7.48}{mm}$). Each frame shows the pattern after $T$ periods. Different stages of the formation can be distinguished, namely (a-c) the initial formation of chains, (c-e) the expansion of the pattern, and (e-f) the contraction towards an equilibrium spacing. A stroboscopic video is provided as Supplemental Material \citep{supmat}.}
    \label{fig:resulttransient}
\end{figure}

We present a typical example of the chain formation in Fig.~\ref{fig:resulttransient}, which corresponds to an experiment in series A1 (see Tab.~\ref{tab:measurementseries}), with $f=\SI{5.0}{Hz}$ and $A\approx\SI{7.48}{mm}$. Additionally, a video is available as Supplemental Material \citep{supmat}. 
At the start of the experiment, the particles are clustered near the walls of the box [Fig.~\ref{fig:resulttransient}(a)].
Within a few periods of the oscillation, short chains form at the edges of the clusters and quickly self-organize into long chains oriented perpendicular to the oscillation direction [Figs.~\ref{fig:resulttransient}(a-c)]. 
Once formed, the chains repel each other, causing the spacing between the chains to increase and the pattern to fill a large part of the domain [Figs.~\ref{fig:resulttransient}(c-e)].
After the expansion, the streamwise motion of the chains reverses, and the spacing between the chains decreases [Figs~\ref{fig:resulttransient}(e-f)]. After approximately 80 periods, the system reaches an equilibrium state in which the spacing between the chains remains constant over time. In this case, the pattern of chains covers only a portion of the domain. Dynamic behavior is mainly present at the particle level, such as the propagation of defects. 

The evolution of the chains in Fig.~\ref{fig:resulttransient} suggests that, between the chains, there are repulsive interactions at small distances and attractive interactions at large distances. In the absence of attractive interactions, repulsion would cause the chains to spread evenly across the full domain, in contradiction to what is observed in Fig.~\ref{fig:resulttransient}(f). These observations support the hypothesis proposed in previous studies \citep{wunenburger2002periodic,klotsa2009chain} that attractive forces between the chains exist at large distances. In Sec.~\ref{sec:mechanism}, we investigate the underlying physical mechanism that drives the repulsive and attractive interactions, but in the remainder of this section, we first focus on describing the pattern characteristics after reaching an equilibrium. 

\subsection{Streamwise particle motion relative to the fluid}\label{sec:streamwise}
The streamwise motion of the particles can be described in terms of their relative excursion length with respect to the box \citep{wunenburger2002periodic,klotsa2007interaction,overveld2022numerical}. If the forces acting on the particles oscillate harmonically, then the streamwise particle motion is sinusoidal\citep{overveld2022effect}, with the relative particle-fluid excursion length
\begin{equation}\label{eq:relativeamplitude}
    A_r = \sqrt{A^2+A_s^2-2AA_s\cos\left(\phi_s-\phi_0\right)},
\end{equation}
where $A$ and $\phi_0$ describe the motion of the box (see Eq.~\eqref{eq:boxmotion}), $A_s$ is the particle excursion length in the lab frame, and $\phi_s$ the relative phase of the particle oscillation. 
If the particle-bottom friction can be neglected, then $A_r$ is proportional to $A$~~\citep{overveld2022effect}. This proportionality also holds when the particle-bottom friction follows a harmonic oscillation., which is actually the case for all our experiments, as shown in the remainder of this subsection and in Appendix~\ref{sec:appendix}.  

The value of $A_r$ is obtained by simultaneously measuring the motion of the box, to obtain $A$ and $\phi_0$, and the motion of the particles, to obtain $A_s$ and phase $\phi_s$. To avoid including particle-wall interactions (e.g. collisions) in the calculation, the latter two quantities are averaged only over the particles that are at least two diameters away from the side walls of the box. The uncertainty in the averages of these two quantities is typically smaller than 1\%.

\begin{figure*}[ht]
    \includegraphics[width=\linewidth]{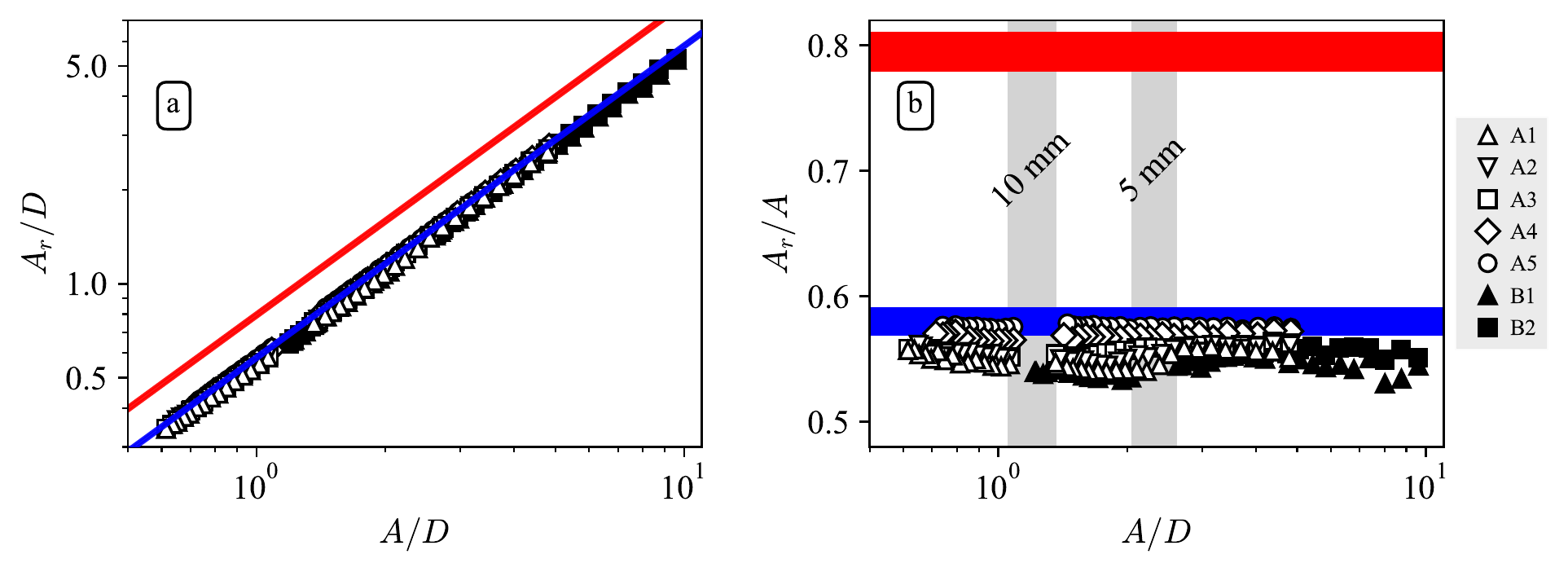}
    \caption{(a) The non-dimensional relative particle-fluid excursion length $A_r/D$ as a function of the non-dimensional excursion length of the box $A/D$. Different symbols correspond to different experiment series, as defined in Table~\ref{tab:measurementseries}. The red and blue regions (in both plots) correspond to Eqs.~\eqref{eq:BBO_frictionless}~and~\eqref{eq:BBO_friction}, which are analytical predictions without and with particle-bottom friction, respectively. (b) The ratio $A_r/A$ is approximately constant ($A_r/A\approx0.55$) for all experiment series and values of $A/D$. The gray shaded areas indicate the gaps in the data (for $D=\SI{10}{mm}$ and \SI{5}{mm}) caused by the resonant shaking of the optical table.}
    \label{fig:relativeamplitude}
\end{figure*}

We calculated $A_r/D$ for all experiments and show the values as a function of $A/D$ in Fig.~\ref{fig:relativeamplitude}(a). The data in this figure collapse onto a line, indicating a unique linear relation between $A_r/D$ and $A/D$. A linear least squares fit is applied to the data, resulting in
\begin{equation}\label{eq:Arfit}
    \frac{A_r}{D} \approx \left(0.552 \pm 0.001 \right)\times \frac{A}{D} + \left(0.009 \pm 0.003 \right),
\end{equation}
which has a high correlation coefficient of $R^2=0.999$. The last term of this equation is small, such that to good approximation $A_r/D\propto A/D$. As further confirmation, Fig.~\ref{fig:relativeamplitude}(b) shows that $A_r/A$ is approximately constant. 

To validate our results, we compare them to the solutions of the Basset-Boussinesq-Oseen (BBO) equation\citep{corrsin1956equation} for the streamwise motion of a single, isolated particle. According to the BBO equation, the ratio $A_r/A$ depends on the particle-fluid density ratio $s$ and the normalized viscous length scale $\delta/D$. The solution for the case without particle-bottom friction is previously determined by \citet{overveld2022effect} and is given by 
\begin{equation}\label{eq:BBO_frictionless}
    \frac{A_r}{A} = \frac{2(s-1)}{\sqrt{(9\delta/D)^2(2\delta/D+1)^2+(9\delta/D+2s+1)^2}}.
\end{equation}
For our experimental conditions with $0.024<\delta/D<0.082$ and $s=7.7$, this solution predicts values for $A_r/A$ in the range $0.78-0.81$. This prediction is indicated by the red bands in Fig~\ref{fig:relativeamplitude}, which do not match our experimental data well.

Next, we consider the case where particle-bottom friction is sufficiently large to cause the particle to roll without slipping, by adding a friction force $\boldsymbol{F}_c$ to the BBO equation. In Appendix~\ref{sec:appendix}, we further explore the role of particle-bottom friction on the streamwise particle motion.
Under the assumption that the particle rotation is only due to particle-bottom friction, the angular momentum of the particle is described by 
\begin{equation}\label{eq:rotationfrictionforce}
    \frac{\rho_s\pi D^5}{60}\frac{d\boldsymbol{\omega}_s}{dt} = \boldsymbol{r}\times\boldsymbol{F}_c,
\end{equation}
where the angular velocity is given by
\begin{equation}\label{eq:angularvelocity}
    \boldsymbol{\omega}_s = \frac{2}{D}(\boldsymbol{u_s}-\boldsymbol{u}),
\end{equation}
with $\boldsymbol{u_s}$ and $\boldsymbol{u}$ being the particle and bottom velocities, respectively.
Equation~\eqref{eq:angularvelocity} is defined such that the particle velocity at the point of contact with the bottom is always equal to the bottom velocity. Following the approach of \citet{overveld2022effect} and incorporating the particle-bottom friction, the solution to the BBO equation is given by
\begin{equation}\label{eq:BBO_friction}
    \frac{A_r}{A} = \frac{2(s-1)}{\sqrt{(9\delta/D)^2(2\delta/D+1)^2+(9\delta/D+14s/5+1)^2}}.
\end{equation}
This solution predicts values of $A_r/A$ in the range $0.57-0.59$, represented by the blue bands in Fig~\ref{fig:relativeamplitude}. These bands describe well our experimental data and agree with the linear fit presented in Eq.~\eqref{eq:Arfit}. 

According to Eq.~\eqref{eq:BBO_friction}, an increase in $\delta/D$ causes a decrease in $A_r/A$. However, the effect is small if $\delta/D\ll s$, which is evident from the narrow blue band in Fig.~\ref{fig:relativeamplitude}(b).
Our experimental data agree with this prediction, as the filled symbols in Fig.~\ref{fig:relativeamplitude}(b) have slightly lower $A_r/A$ values compared to the empty symbols. The spread in the experimental data is only a few percent, consistent with the width of the blue band. Therefore, we conclude that variations in $\delta/D$ do not significantly affect the value of $A_r/A$ in this part of the parameter space.

The solutions to the BBO equation imply that the particles in our experiments roll without slipping, which agrees with our visual observations. Moreover, the collapse of the data onto a single line in Fig.~\ref{fig:relativeamplitude}(a) and the constant value of $A_r/A$ suggest that the relative importance of the particle-bottom friction compared to the driving forces is constant. All in all, the results confirm that the small variations in $\Gamma$ observed in our experiments have no significant effect on the streamwise particle motion.

The values of $A/D$ in Fig.~\ref{fig:relativeamplitude} are not evenly spaced along the horizontal axis. There is a larger gap between data points near the resonance frequency of the optical table (around $\SI{4.3}{Hz}$), corresponding to $A\approx\SI{10}{mm}$, as indicated by the gray shaded areas in Fig.~\ref{fig:relativeamplitude}(b). The shaking of the optical table at these frequency values, as discussed in Sec.~\ref{sec:experimentalsetup}, results in values of $A/D$ that are significantly higher than the user-defined amplitude for the motor.

\subsection{Mean pattern characteristics}\label{sec:patterncharacteristics}
\subsubsection{Qualitative comparison}
\begin{figure*}[ht]
    \includegraphics[width=\linewidth]{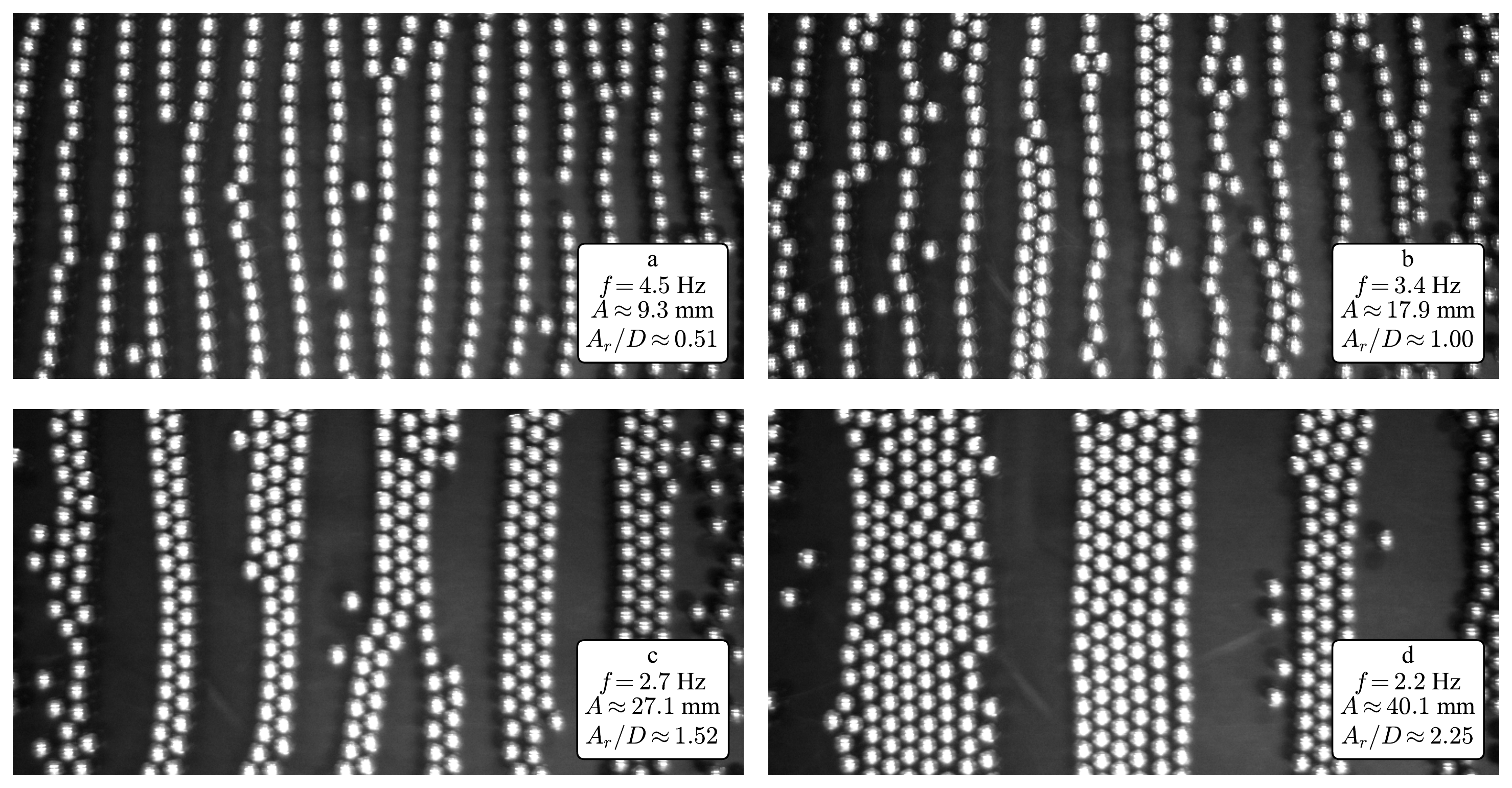}
    \caption{Typical examples of the patterns in experiment series A3 ($D=\SI{10}{mm}$, $\phi\approx0.346$) under different oscillatory conditions. At low values of $A_r/D$, the particles self-organize into one-particle-thick chains, oriented perpendicular to the direction of oscillation (a). As $A_r/D$ increases, a combination of chains and two-particle-wide bands is observed (b). Further increasing $A_r/D$ results in the formation of broader bands that are three (c) or more (d) particles wide. A video is available as Supplemental Material \citep{supmat}.}
    \label{fig:resultoverview}
\end{figure*}

After starting the oscillatory motion of the box, the particles arrange themselves into different types of patterns, depending on the value of $A_r/D$. 
We first consider a single coverage fraction $\phi$ and describe the pattern characteristics for various oscillatory conditions. A comprehensive overview of the parameter space, including variations in $\phi$, is given in Sec.~\ref{sec:phasespace}.
Typical examples of these patterns for the experiment series A3 are shown in Fig.~\ref{fig:resultoverview}, and a video is available as Supplemental Material \citep{supmat}. 
For low values of $A_r/D$, the particles form one-particle-thick chains oriented perpendicular to the direction of oscillation, as shown in Fig.~\ref{fig:resultoverview}(a) and also in the previous example in Fig.~\ref{fig:resulttransient}. This pattern has been described in previous studies by \citet{wunenburger2002periodic} and \citet{klotsa2009chain}.
At slightly larger values of $A_r/D$, the pattern consists of particle chains that are more tortuous, have more defects, and have sections that are two particles wide, as seen in Fig.~\ref{fig:resultoverview}(b).
As $A_r/D$ is further increased, one-particle-thick chains become less prevalent and bands of two and three particles wide become dominant, as shown in Fig.~\ref{fig:resultoverview}(c). 
This trend continues as the particles form wider bands for larger values of $A_r/D$, as shown in Fig.~\ref{fig:resultoverview}(d).
Furthermore, it is common for bands with different widths to coexist in this part of the parameter space. For example, in Fig.~\ref{fig:resultoverview}(d), bands with approximately seven, six, and three rows can be observed simultaneously.

Due to the chaotic nature of many-particle systems, exactly reproducing the patterns from our experiments is impossible. Small variations in the initial conditions lead to significant differences in the particle trajectories. However, if we repeat our experiments at the same conditions, we obtain statistically equivalent patterns. Notably, the mean characteristics of the patterns once they have reached an equilibrium state are similar. 

Additionally, we frequently observe collisions between particles and the walls that are perpendicular to the oscillating direction, as can be clearly seen in the video available as Supplemental Material \citep{supmat}. Due to such a collision with the moving wall, particles rebound toward the center of the box with a significantly increased velocity. Subsequently, these isolated particles often collide with particle chains or bands, leading to defects in the pattern near these walls.  
However, to maintain a constant particle coverage and avoid arbitrary cutoff points, we include all particles in the box in the analysis of the mean pattern characteristics.

We characterize the mean state of the system using a radial pair correlation function (RPCF), $g_r$, which is commonly used for colloidal systems \citep{lotito2020pattern}. The RPCF describes the particle number density around a reference particle, as a function of the distance to its center. It is defined as
\begin{equation}
    g_r = \frac{1}{2\pi r\Delta r N\phi}\sum\limits_{i=1}^N\sum\limits_{j=i+1}^N \delta_r^{ij},
\end{equation}
with
\begin{equation}\label{eq:deltar}
    \delta_r^{ij}=
    \begin{cases}
        1, & \text{if } r-\dfrac{\Delta r}{2}<r_{ij}<r+\dfrac{\Delta r}{2},\\
        0, & \text{otherwise},\\
    \end{cases}
\end{equation}
and where $r$ is the radial coordinate with respect to a reference particle, $\Delta r$ is the bin width, and $r_{ij}$ is the distance between the centers of particles $i$ and $j$.

\begin{figure}[h]
    \includegraphics[width=\linewidth]{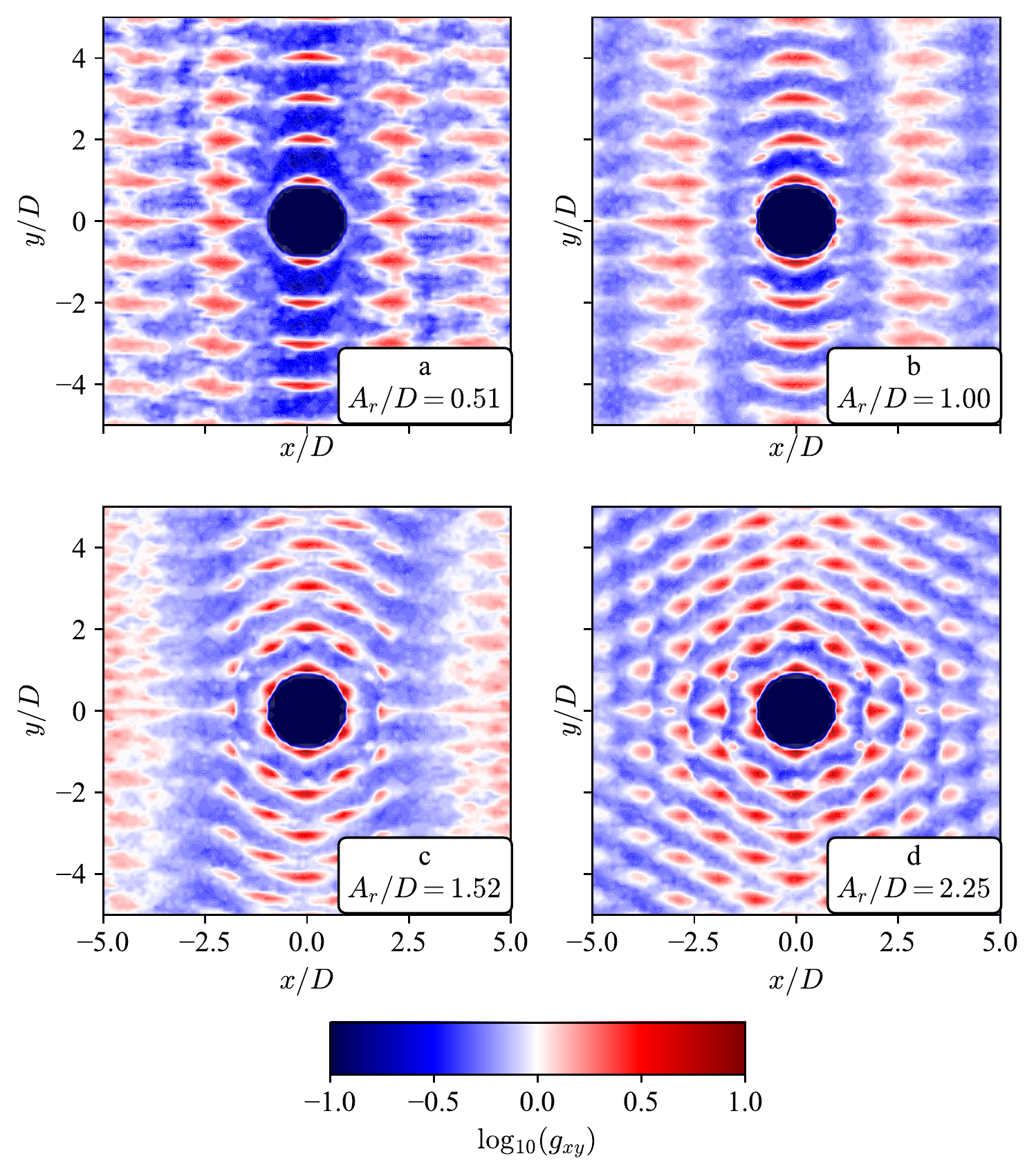}
    \caption{(Color online) The two-dimensional pair correlation function (2DPCF) $g_{xy}$, as defined in Eq.~\eqref{eq:gxy}, is plotted for the four examples (a-d) from Fig.~\ref{fig:resultoverview}, using a bin width of $\Delta x=\Delta y = 0.1D$. The 2DPCF is time-averaged over 100 frames, recorded at different phases of the oscillation. Positive correlations ($\log_{10}(g_{xy})>0$) are indicated in red, negative correlations ($\log_{10}(g_{xy})<0$) in blue, and no correlation ($\log_{10}(g_{xy})=0$) is indicated in white.}
    \label{fig:gxyoverview}
\end{figure}

This approach works best for patterns in isotropic systems since directionality is not considered. To better analyze the anisotropic patterns in our experiments, we use a modified two-dimensional pair correlation function (2DPCF), $g_{xy}$, defined by 
\begin{equation}\label{eq:gxy}
    g_{xy} = \frac{1}{\Delta x \Delta y N\phi} \sum\limits_{i=1}^N\sum\limits_{j=i+1}^N \delta_x^{ij}\delta_y^{ij},
\end{equation}
where $\delta_x^{ij}$ and $\delta_y^{ij}$ have equivalent definitions to $\delta_r^{ij}$ in Eq.~\eqref{eq:deltar}. Here, $x$ and $y$ are the Cartesian coordinates parallel and perpendicular to the direction of oscillation, respectively. Along these directions, $\Delta x$ and $\Delta y$ are the bin widths, and $x_{ij}$ and $y_{ij}$ are the distances between the centers of particles $i$ and $j$ in the $x$- and $y$-direction, respectively.
We use Eq.~\eqref{eq:gxy} to analyze the four cases shown in Fig.~\ref{fig:resultoverview}. The results are averaged over 100 frames per experiment, taken at different phases of the oscillation. The result is shown in Fig.~\ref{fig:gxyoverview} on a logarithmic scale.
We evaluate the function $g_{xy}$ solely at the center of each bin, with width $\Delta x=\Delta y = 0.1D$, such that particles are not double-counted.

For all four cases, the 2DPCF shown in Fig.~\ref{fig:gxyoverview} shows a distinct structure that is qualitatively similar to the patterns observed in the experiments. The red regions indicate a higher-than-average particle presence. The circular region around the origin represents the reference particle, which does not overlap with other particles, such that $g_{xy}=0$. 
For low values of $A_r/D$, as shown in Fig.\ref{fig:gxyoverview}(a), the red regions in the 2DPCF represent one-particle-thick chains, with particles within each chain appearing at $y/D=\pm1,\pm2,\dots$. The red regions are elongated in the $x$-direction due to the tortuosity of the chains.
Neighboring chains (or bands) can be distinguished at $x/D\approx\pm 2$ and $\pm 4$ in Fig.~\ref{fig:gxyoverview}(a), at $x/D\approx\pm3$ in Fig.~\ref{fig:gxyoverview}(b), and at $x/D\approx\pm5$ in Fig.~\ref{fig:gxyoverview}(c). The spacing between neighboring chains or bands thus clearly increases with $A_r/D$, which we address in more detail in Sec.~\ref{sec:distance}.

Furthermore, the neighboring rows in Figs.~\ref{fig:gxyoverview}(a)~and~(b) have peaks at $y/D=\pm1,\pm2,\dots$, similar to the central row, indicating that the particles in neighboring chains are aligned. This is likely a result of confinement: when two neighboring chains both touch a side wall, the particles closest to the wall must be aligned. This alignment might diminish further away from the walls, due to e.g. tortuosity and defects. However, our domain is not sufficiently large to discard the influence of the side walls.
As $A_r/D$ increases and the width of the bands increases, more regions with $g_{xy}>1$ around the central row appear. Due to the point symmetry of $g_{xy}$ about the origin, these regions are present on both sides of the central row and correspond to a hexagonal packing, which is most evident in Figs.~\ref{fig:gxyoverview}(c)~and~(d).
Overall, the peak values of $g_{xy}$ decrease with distance from the origin due to the finite size of the domain. Additionally, the patterns in the 2DPCF may be slightly skewed, such as in Fig.~\ref{fig:gxyoverview}(b). This skewness is due to the chains not being perfectly perpendicular to the direction of oscillation, which can be caused by small variations in the horizontality and smoothness of the bottom plate.

\subsubsection{Distance between chains}\label{sec:distance}
\begin{figure}[h]
    \includegraphics[width=\linewidth]{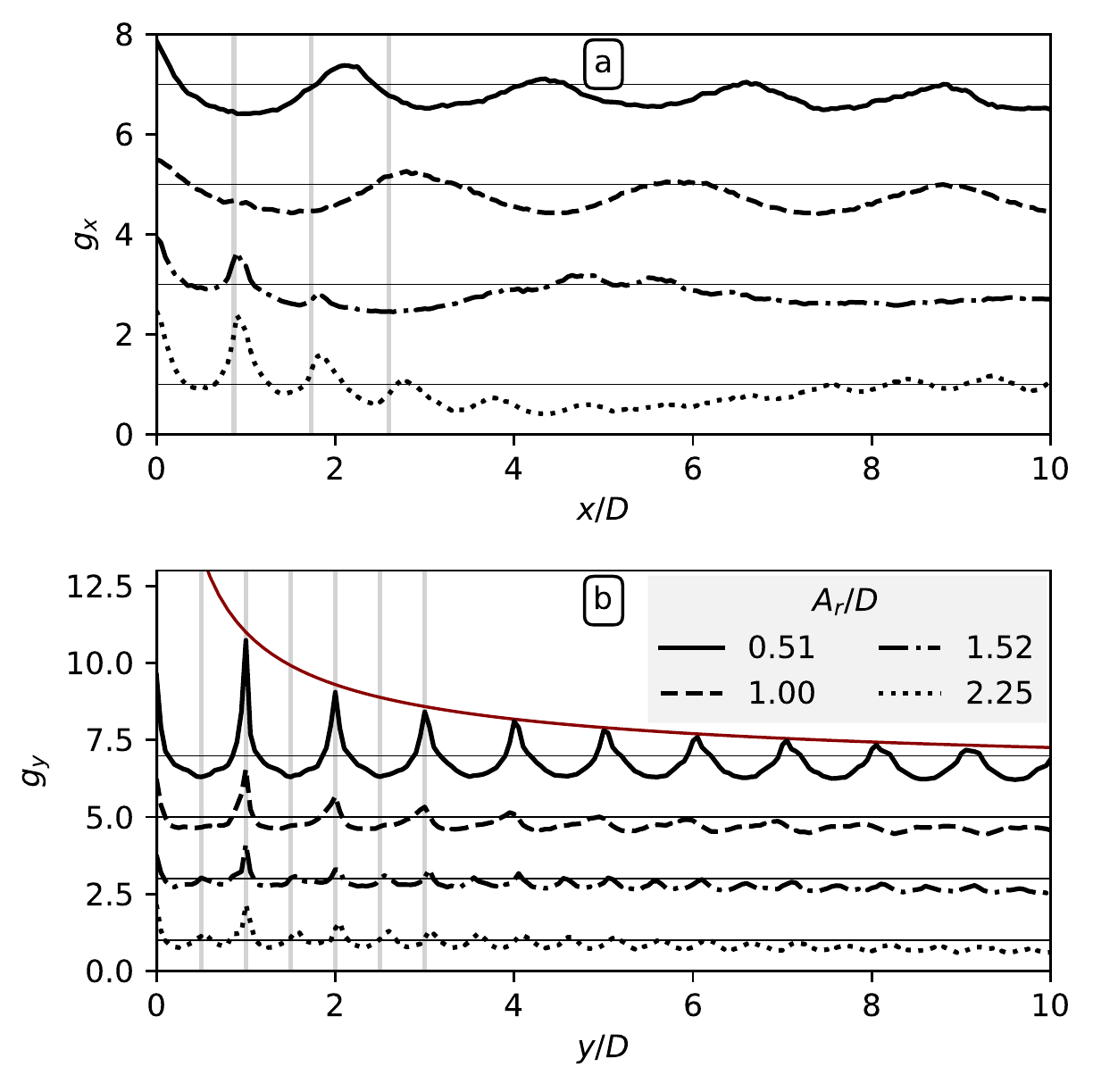}
    \caption{The one-dimensional pair correlation functions (a) $g_x$ and (b) $g_y$ are calculated for the four cases shown in Fig.~\ref{fig:resultoverview}. We have used Eq.~\eqref{eq:gxy} with ($\Delta x=0.01, \Delta y=10$) and ($\Delta x=10, \Delta y=0.01$) for $g_x$ and $g_y$, respectively. The curves have been shifted by multiples of two for improved clarity.
    The horizontal lines are the reference values for which no correlation is present ($g_x$ or $g_y$ equal to 1).
    The vertical lines in (a) correspond to multiples of $\sqrt{3}/2$, whereas the vertical lines in (b) correspond to multiples of $1/2$. The red curve in (b) represents the envelope $g_y \sim (y/D)^{-0.6}$).}
    \label{fig:gxycollapse}
\end{figure}

The two-dimensional representations in Fig.~\ref{fig:gxyoverview} are useful for identifying qualitative differences between the average patterns, but extracting quantitative information directly from them is challenging. 
Instead, we use one-dimensional pair correlation functions $g_x$ and $g_y$ for the $x$- and $y$-directions, respectively, to obtain characteristic length scales from the pattern. 
For $g_x$, we set $\Delta y$ to a large value ($\Delta y = 10D$), while keeping $\Delta x$ small ($\Delta x=0.01D$) to obtain a high spatial resolution in the $x$-direction. Similarly, for $g_y$, we set $\Delta x=10D$ and $\Delta y = 0.01D$. 
These operations are equivalent to averaging $g_{xy}$ in either the $y$- or $x$-direction, respectively. The results are shown in Fig.~\ref{fig:gxycollapse}, for the same four cases of series A3 displayed in Figs.~\ref{fig:resultoverview}~and~\ref{fig:gxyoverview}.

For low values of $A_r/D$ ($A_r/D=0.51$, 1.00), represented by the solid and dashed curves in Fig.~\ref{fig:gxycollapse}, the peaks in $g_x$ correspond to the neighboring chains. The width of these peaks, which is on the order of one diameter, is primarily caused by variations in the chains, such as defects and tortuosity. The peak widths are typically more than one order of magnitude larger than the measurement uncertainty, which is on the sub-millimeter level (see Sec.~\ref{sec:experimentalsetup}).
Between the peaks, the value of $g_x$ is significantly smaller than one, indicating that there is a lower-than-average chance of finding a particle in that region. 
Contrarily, in $g_y$, we observe sharp peaks at integer values of $y/D$, which correspond to the particles within the chains. The peaks decay as the distance from the reference particle increases, as indicated by the red envelope in Fig.~\ref{fig:gxycollapse}(b). The observed algebraic decay is a characteristic feature of quasi-long-range translational order in crystalline structures \citep{lotito2020pattern}. Additionally, the finite size of the box in our system also leads to a less frequent occurrence of particles at large distances.

At higher values of $A_r/D$ ($A_r/D=1.52$, 2.25), represented by the dash-dotted and dotted curves in Fig.~\ref{fig:gxycollapse}, additional peaks are visible in both $g_x$ and $g_y$. For $g_x$, these appear approximately at $(1/2)\sqrt{3}$, $\sqrt{3}$, and $(3/2)\sqrt{3}$, corresponding to the distances in the $x$-direction between spheres in a hexagonal packing. The peak values are slightly shifted towards larger values, which can be attributed to the outermost rows of particles in the bands not always being tightly packed, as can be observed in Fig.~\ref{fig:resultoverview}(d).
The broad peaks in the range $4<x/D<6$ (dash-dotted) or $7<x/D<10$ (dotted) correspond to the neighboring bands. Additionally, the smaller peaks modulating the broad peaks correspond to individual rows within these bands.
In $g_y$, the peaks at integer values diminish with increasing $A_r/D$. Additional peaks appear at half-integer values ($0.5,1.5,2.5\dots$), which correspond to the distances in a hexagonal arrangement.

\begin{figure}[h]
    \includegraphics[width=\linewidth]{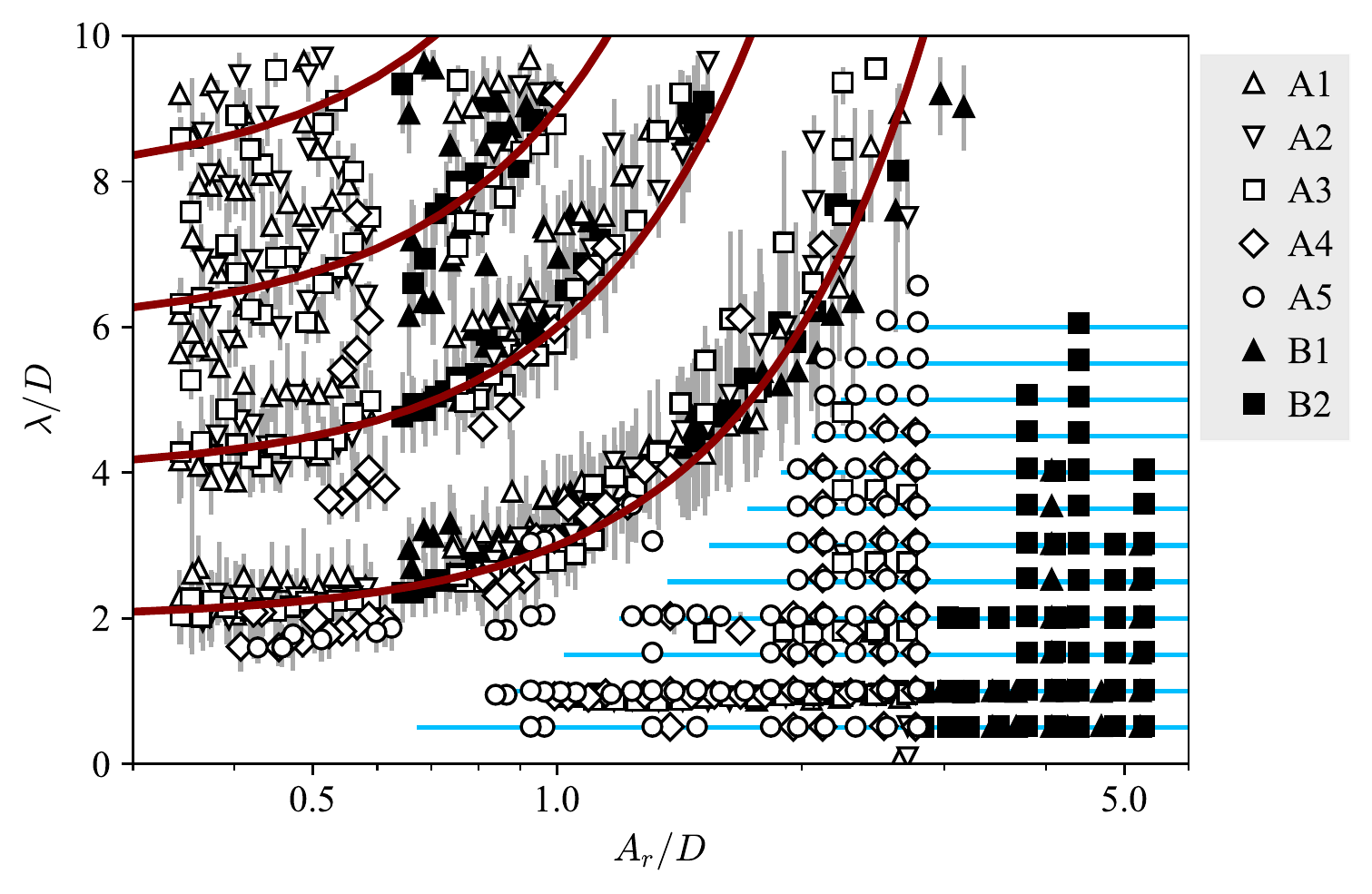}
    \caption{(Color online) The positions of the peaks in $g_x$ as a function of $A_r/D$, for all experiment series (see Tab.~\ref{tab:measurementseries}). The error bars represent the full width at half maximum (FWHM) of these peaks. The (red) curves are integer multiples of Eq.~\eqref{eq:chainspacing}. The (blue) horizontal lines correspond to integer multiples of $\sqrt{3}/2$.}
    \label{fig:chainspacing}
\end{figure}

To quantify the distance between the chains, we calculate the (normalized) interchain distance $\lambda/D$ and its full width at half maximum (FWHM) of each peak in $g_x$ for all experiments in each experiment series. The results are plotted as a function of $A_r/D$ in Fig.~\ref{fig:chainspacing}, with the error bars representing the FWHM values. 
The data from the different experiment series collapse onto several curves without requiring rescaling. The data on the lowest curve, starting at $\lambda/D\approx2$, corresponds to the neighboring chains or bands, while the data with larger values of $\lambda/D$ correspond to non-neighboring chains or bands.

We have excluded a part of the oscillating box when calculating $g_x$ for the experiments from series A5 with $A_r/D<0.7$. This is necessary because irregularly shaped clusters tend to form during these experiments, particularly close to the side walls. The peaks in $g_x$ associated with these clusters would have obscured the data due to the chain spacing, making it difficult to analyze the results accurately. In Sec.~\ref{sec:phasespace} we address the origin of these clusters in this part of the parameter space.

Based on the data from the lowest curve in Fig.~\ref{fig:chainspacing}, we observe that the distance between the centers of neighboring chains follows 
\begin{equation}\label{eq:chainspacing}
    \frac{\lambda}{D} \approx 2 + \left(\frac{A_r}{D}\right)^2.
\end{equation}
The red curves in the figure represent the value of $\lambda/D$ and its integer multiples (i.e. $2\lambda/D$, $3\lambda/D$ and $4\lambda/D$) as given by Eq.~\eqref{eq:chainspacing}. This equation further implies a minimum spacing between neighboring chains equal to one particle diameter ($\lambda=2D$) when $A_r/D$ approaches zero. Moreover, the spacing between neighboring chains is equal to two particle diameters ($\lambda=3D$) when $A_r/D=1$.

The experimental data in Fig.~\ref{fig:chainspacing} shows good agreement with the proposed relation for the chain spacing (lowest red curve, corresponding to Eq.~\eqref{eq:chainspacing}) across all values of the particle coverage fraction $\phi$. Even for low values of $\phi$ where the pattern does not fully cover the box, the data follows the curve. 
The chain spacing is thus primarily determined by hydrodynamic interactions between the chains, which are characterized by $A_r/D$. Equation~\eqref{eq:chainspacing} further implies that the spacing is set by the equilibrium between short-range repulsion and long-range attraction. Conversely, if the chain solely repel each other, the pattern would spread out over the box and the spacing would be determined by the confinement, i.e. by $\phi$.

However, we note that for $A_r/D\lesssim0.6$ in Fig.~\ref{fig:chainspacing}, the values of $\lambda/D$ systematically decrease as $\phi$ increases. Specifically, the data points for small values of $\phi$ (e.g the triangles) lie slightly above the red curve, while the data points for large values of $\phi$ (e.g. the diamonds and the circles) lie slightly below the red curve. This result implies that confinement does play a role in this part of the parameter space, which we further discuss in the context of the interaction potentials in Sec.~\ref{sec:phasespace}.

Our results are further consistent with those of \citet{wunenburger2002periodic}, who reported that the chain spacing is independent of the particle coverage fraction $\phi$ for low values (around $\phi\approx 0.07$). We confirm this finding and extend it to higher values of $\phi$ by almost one order of magnitude.
Nonetheless, the scaling proposed by \citet{wunenburger2002periodic} differs considerably from ours, as they suggested that $\lambda/D\sim \left(A_r/D\right)^{0.5}\left(A_r(2\pi f) D/\nu\right)^{-0.21}$. This can be rewritten to $\lambda/D\sim\left(A_r/D\right)^{0.29}\left(\delta/D\right)^{0.42}$, implying that the spacing depends on both $A_r/D$ and $\delta/D$. 
However, for our experiments, we do not find such a dependence on $\delta/D$ and we can show that the relation proposed by \citet{wunenburger2002periodic} does not hold for our results. For constant values of $\Gamma$ and $A_r/D$, the ratio $\delta/D$ is a function of the particle size: $\delta/D\sim\left(gD^3/\nu^2\right)^{-1/4}$, where the term between brackets is commonly known as the Galilei number. Therefore, experiments with identical values of $\Gamma$ and $A_r/D$ but different particle diameters result in different values of $\delta/D$. In Fig.~\ref{fig:chainspacing}, we observe that different symbols overlap at a constant value of $A_r/D$, regardless of variations in $\delta/D$ through different frequencies and particle sizes. Hence, we conclude that within the region of the parameter space we explored, $A_r/D$ is a relevant parameter for the chain spacing, but $\delta/D$ is not.

In the bottom-right corner of Fig.~\ref{fig:chainspacing}, for large $A_r/D$, the data represent the spacing between particles within the same band with hexagonal packing. The FHWM is small in this region, which indicates that the spacing between individual rows within a band deviates little from the peak positions. 
For $D=\SI{10}{mm}$ (open symbols), the values are integer multiples of $\sqrt{3}/2$ (indicated by the blue horizontal lines), as expected from the peak positions in Fig.~\ref{fig:gxycollapse}. However, for $D=\SI{5}{mm}$ (filled symbols), some peaks are located in between the blue lines, at integer multiples of $1/2$. To understand the differences, we compare the patterns for different particle sizes. 

\begin{figure*}[ht]
    \includegraphics[width=\linewidth]{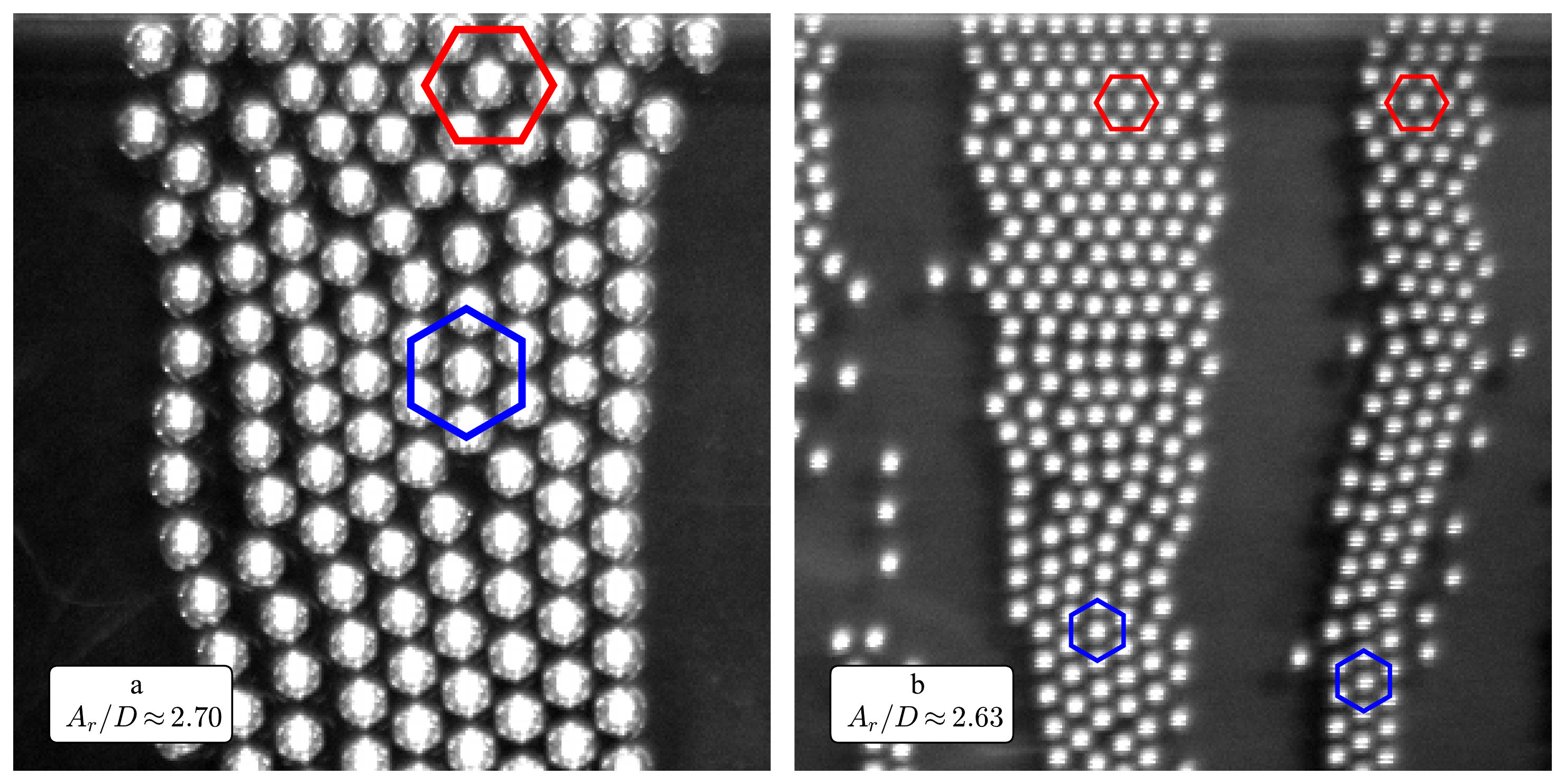}
    \caption{(Color online) Images from two experiments with (a) $D=\SI{10}{mm}$ and (b) $D=\SI{5}{mm}$, for similar values of $A_r/D$ and plotted on the same scale. The particles in each band are closely packed in a hexagonal arrangement with different orientations in different parts of the domain. Near the side wall (top), the orientation is such that the particles align with it, whereas far away from the boundaries (bottom), the orientation is rotated by $30^\circ$. The hexagons are added to clarify the orientations near the wall (red) and in the bulk (blue).}
    \label{fig:patternshapes}
\end{figure*}

The hexagonal arrangement within each band has two distinct orientations, rotated $30^\circ$ with respect to each other, as shown in Fig.~\ref{fig:patternshapes}. 
Close to the side walls, the particles form rows that are aligned with and touching the walls. This suggests that particles are attracted to the walls, and that this attraction is stronger than the attraction to neighboring particles on the other side. Likewise, \citet{klotsa2009chain} observed that one-particle-thick chains are attracted to the side walls.
Further away from the boundaries, each band instead consists of a staggered arrangement of particle chains, where each chain is oriented perpendicular to the oscillation direction.
Between the two regions with different orientations, there is a region of a few particles wide with many defects.
The region with wall-oriented particles is significantly narrower for the $\SI{10}{mm}$ particles than for the $D=\SI{5}{mm}$ particles, with the region being approximately 4 and 13 particles wide for the specific cases in Figs.~\ref{fig:patternshapes}(a)~and~(b), respectively. There are thus significantly more particles in the wall-oriented part for the smaller particles, which leads to the peaks at integer multiples of $1/2$ in $g_x$ in Fig.~\ref{fig:chainspacing}.

\subsubsection{Chain width}\label{sec:chainwidth}
To describe the chain width, we use the fraction
\begin{equation}\label{eq:Fn}
    F_n = \frac{M_n}{N},
\end{equation}
where $M_n$ is the number of particles with $n$ nearest neighbors and $N$ is the total number of particles. Two particles are considered nearest neighbors if their center-to-center distance is smaller than a cutoff value: $r_{ij}/D<1.3$. This cutoff value is chosen based on previous work on periodic patterns, which suggests that a value should be chosen between the closest and the next-closest particle distance in the packing (between 1 and $\sqrt{2}$ for square packing and between 1 and $\sqrt{3}$ for hexagonal packing) \citep{lotito2020pattern,besseling2012oscillatory}.

\begin{figure*}[ht]
    \includegraphics[width=\linewidth]{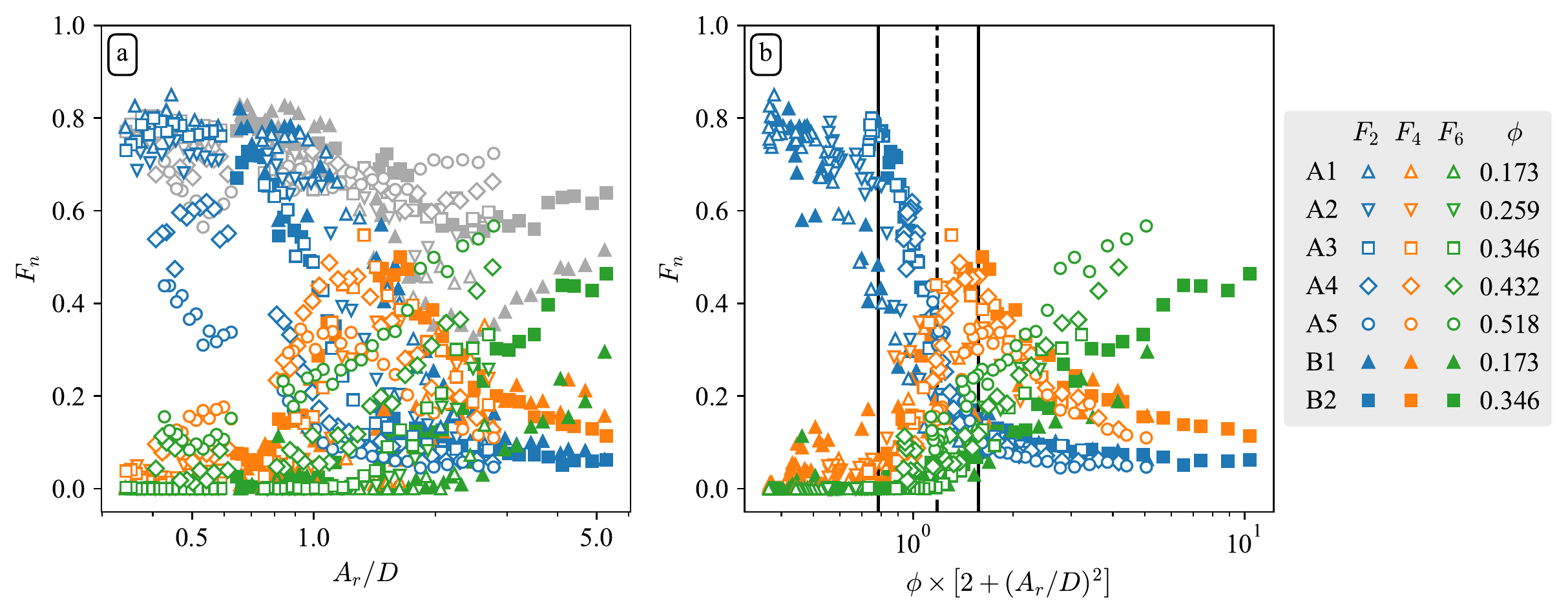}
    \caption{(Color online) The fraction of particles $F_n$ with $n$ nearest neighbors (see Eq.~\eqref{eq:Fn}) as a function of (a) $A_r/D$ and (b) $\phi\times \left[2+\left(A_r/D\right)^2\right]$, for all experiment series (see Tab.~\ref{tab:measurementseries}). The gray symbols in (a) represent $F_2+F_4+F_6$, i.e. the fraction of particles that is not a defect. The vertical lines in (b) correspond to the critical values $\pi/4$ and $\pi/2$, according to Eqs.~\eqref{eq:ArDcrit}~and~\eqref{eq:ArDcrit2}, respectively. The dotted line at $\phi[2+(A_r/D)^2]=3\pi/8$ lies halfway between the two critical values.}
    \label{fig:chainwidth}
\end{figure*}

In Fig.~\ref{fig:chainwidth}(a), we present the fractions of particles with 2, 4, and 6 nearest neighbors as a function of $A_r/D$. Across all experiment series, a similar trend is observed which matches observations from the experiments, e.g. as shown in Fig.~\ref{fig:resultoverview}. At low $A_r/D$, $F_2\approx0.75$, meaning that about 70-80\% of particles are part of one-particle-thick chains where each particle has two nearest neighbors.
As $A_r/D$ increases, $F_2$ decreases while $F_4$ increases, signifying that particles form bands that are two rows wide.
The increase of $F_6$ at even higher values of $A_r/D$ shows that the bands become wider, with at least three rows.
Additionally, the symbols from experiment series with equal values of $\phi$ fall on top of each other. For example, open and filled squares overlap, and open and filled upward triangles overlap. Conversely, when $\phi$ is increased, the transitions of $F_2$ to $F_4$ and $F_4$ to $F_6$, occur at lower values of $A_r/D$. The particles not accounted for are those with zero, one, three, and five neighbors that are related to isolated particles and defects in the chains and bands. For each experiment, $\sum_{n=0}^6 F_n=1$.   

For the cases with large $\phi$ (open circles and open diamonds) and $A_r/D\lesssim0.6$, approximately 20\% of the particles have either four or six nearest neighbors. These particles are part of irregularly shaped clusters that emerge when $\phi$ is sufficiently large. In Sec.~\ref{sec:phasespace}, we provide an explanation for why these clusters are only found in these particular experiment series.

The gray symbols in Fig.~\ref{fig:chainwidth}(a) represent the sum $F_2+F_4+F_6$, which is the fraction of particles that are neither defects nor isolated particles. At low values of $A_r/D$, around $20\%$ of the particles are part of defects or isolated, and this percentage increases to $40-60\%$ for larger values of $A_r/D$. 
Notably, the cases with low particle coverage fractions tend to have more defects for large $A_r/D$. For instance, the upward triangles in Fig.~\ref{fig:chainwidth}(a) demonstrate that around $A_r/D\approx2$, only 30-40\% of the particles have two, four, or six nearest neighbors. Furthermore, we observe that in these cases, there is no peak in $F_4$, in contrast to the well-defined peaks for slightly higher $\phi$ values (squares, diamonds, and circles).

We propose a model to explain the observed variations in chain widths by connecting them to earlier observations on the distance between the chains. 
We first assume that all particles are arranged in chains spanning between both side walls, i.e. with length $L_y$ equal to the width of the domain. Each chain contains $L_y/D$ particles and there are a total of $ND/L_y$ chains in the system. Next, we assume that the distance between the chains, as given by Eq.~\eqref{eq:chainspacing}, is an intrinsic length scale that is independent of the confinement. 
The collection of chains then spans a distance 
\begin{equation}
    L_c=\frac{ND^2}{L_y}\left[2+\left(\frac{A_r}{D}\right)^2\right]
\end{equation} 
in the $x$-direction, where edge effects have been neglected. The collection fits inside the domain when $L_c<L_x$, with $L_x$ the length of the domain. Hence, two-particle-wide bands should form when $L_c>L_x$, in other words, when
\begin{equation}\label{eq:ArDcrit}
    \phi\left[2+\left(\frac{A_r}{D}\right)^2\right]>\frac{\pi}{4},
\end{equation}
where we have used the definition of the particle coverage fraction from Eq.~\eqref{eq:coveragefraction} ($\phi=\pi N D^2/(4L_xL_y)$).
The condition in Eq.~\eqref{eq:ArDcrit} is satisfied for all values of $A_r/D$ if $\phi>\pi/8\approx0.39$. Above this critical particle coverage fraction, the spacing between one-particle-thick chains is always incompatible with the domain size and the particles should always form bands to accommodate the preferred spacing.

The previous analysis can be repeated for the case of two-particle-wide bands, where each band contains $2L_y/D$ particles, while the expression for the spacing remains unchanged. This analysis yields that three-particle-wide bands should form when
\begin{equation}\label{eq:ArDcrit2}
    \phi\left[2+\left(\frac{A_r}{D}\right)^2\right]>\frac{\pi}{2}. 
\end{equation}
It is worth noting that Eqs.~\eqref{eq:ArDcrit}~and~\eqref{eq:ArDcrit2} depend on $\phi$ and not explicitly on the domain size. Thus, even though the model is derived for a bounded domain, it is also valid for periodic and infinitely large domains, where the intrinsic spacing between the chains may also become incompatible with the coverage fraction.

We validate our model by plotting the fraction of particles with $n$ nearest neighbors, $F_n$, as a function of the left-hand side of Eq.~\eqref{eq:ArDcrit} in Fig.~\ref{fig:chainwidth}(b). Here, we observe the collapse of the data from different values of $\phi$ from Fig.~\ref{fig:chainwidth}(a).
Furthermore, at the critical value $\pi/4$, there is a decrease in $F_2$ and an increase in $F_4$ marking the transition from one-particle-thick chains to two-particle-wide bands. For higher coverage fractions, i.e. the circles, diamonds, and squares, the transition is sharper than for lower coverage fractions, i.e. the triangles. 
Moreover, the critical value $\pi/2$ obtained from Eq.~\eqref{eq:ArDcrit2} marks the maximum in $F_4$ and the onset of the increase in $F_6$. These two changes in $F_n$ correspond to the transition from two- to three-particle-wide bands.
Finally, note that the blue and orange symbols in Fig.~\ref{fig:chainwidth}(b) intersect approximately halfway between the two transitions, at the dotted line.

\section{\label{sec:mechanism}Physical mechanism for chain attraction and repulsion}
\subsection{Numerical method}
We uncover the physical mechanism behind the attraction and repulsion between the chains using numerical simulations. We employ a second-order accurate finite volume method that fully resolves the flow around the spherical particles. The particles are simulated using an immersed boundary method, as implemented by \citet{breugem2012second}.
This code has been recently used to study the steady streaming flows and dynamics of particle pairs in oscillating flows by \citet{overveld2022numerical,overveld2022effect}. For a detailed explanation of the numerical method and its adaptation for oscillating flows, we refer the reader to the aforementioned sources. 

In our simulations, we are primarily interested in the time-averaged flow around the chains for different configurations and flow conditions. To allow for an easier comparison, the particles are fixed in space, while the flow and bottom oscillate with a user-defined amplitude. Note that $A_r$ and $A$ are equal in these cases.
Specifically, we simulate cases in which either one or two perfectly straight chains of ten spheres are placed inside a double periodic box with lateral dimensions of $20\times10$ and a height of 5 diameters. The flow is solved on a uniform Cartesian grid with a grid spacing of $D/24$.
We further set the ratio between the viscous length scale and the particle diameter to $\delta/D=1/30$. This corresponds to a frequency of \SI{2.86}{Hz} for particles with \SI{10}{mm} diameter and, given that $\Gamma=0.75$, an amplitude-to-diameter ratio of $A/D\approx2.3$. We average the flow fields over a single period once the transients have sufficiently died out, which is typically after 20 periods. Additionally, the time-averaged flow fields are averaged over the $y$-axis, which yields an average flow field in the $xz$-plane.

\subsection{Steady streaming and resulting forces}
\begin{figure*}[ht]
    \centering
    \includegraphics[width=\textwidth]{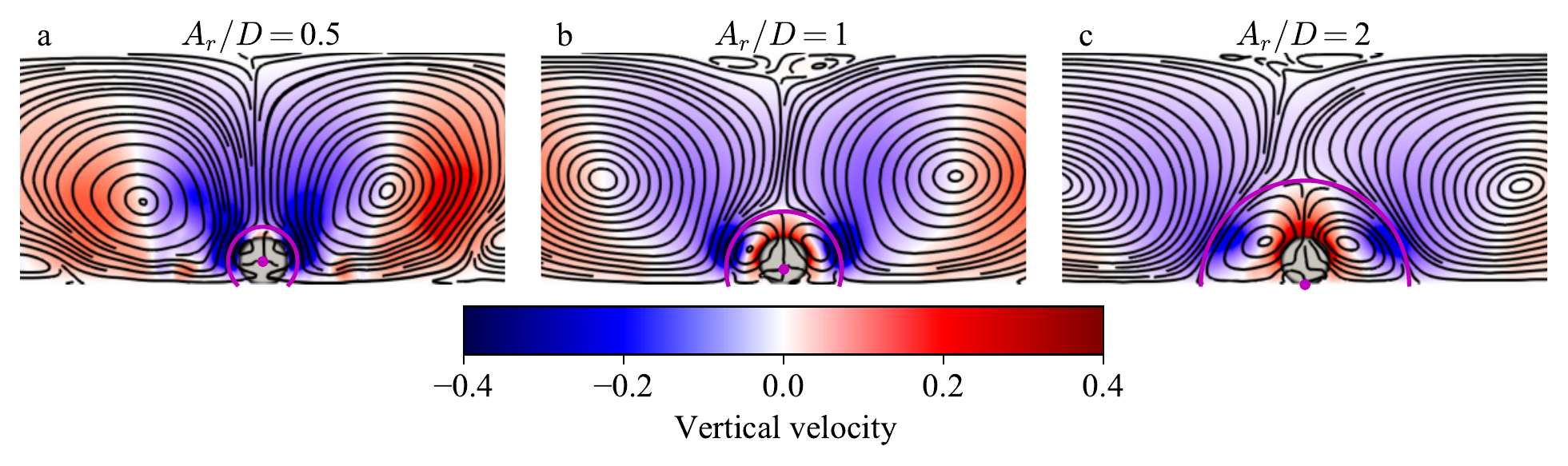}
    \caption{(Color online) The side-view of the flow field around a single particle chain from our numerical simulations, for three different values of $A_r/D$. The flow fields are averaged over a full period of the oscillation and over the $y$-direction. The black lines are streamlines of the average velocity field and the color corresponds to the vertical velocity, normalized by the velocity amplitude $2\pi A f$.
    The sizes of the vortices close to the chain are roughly proportional to $A_r/D$, as indicated by the magenta circles with radii of $[0.75,1.25,2.25]$, respectively. Note that the circle midpoints are not at the same location in the different images, they shift vertically downwards with increasing $A_r/D$.}
    \label{fig:simulationsonechain}
\end{figure*}

Figure~\ref{fig:simulationsonechain} shows the flow field around a single chain of particles for three different values of $A_r/D(=A/D)$. The simulations reveal the presence of small `inner' vortices near the particles and larger `outer' vortices that fill most of the domain and span the full height.
In this two-dimensional representation, the inner vortices appear to partially overlap with the particles, especially for $A_r/D=0.5$. The vortices are actually located in between successive particles, i.e. at different $y$-positions compared to the particle centers.
Additionally, combining the vertical velocity and streamlines, we deduce that a downward flow above the chain (blue) signifies a diverging flow in the horizontal $xy$-plane at mid-particle height. Conversely, an upward flow above the chain (red) signifies a converging flow in this plane.

The radius of the inner vortices is roughly proportional to $A_r/D$, as indicated by the magenta circular arcs in the figure. The size difference of these vortices leads to qualitative differences in the flow around the chains. For $A_r/D=0.5$, the inner vortices are smaller than a particle diameter, leading to a downward flow above the chain and flow away from the chain at mid-particle height. In contrast, for the larger values of $A_r/D$ (1.0 and 2.0), the inner vortices are substantially larger than one particle diameter and are mainly located next to the projection of the chain. This results in a flow directed towards the particle at mid-particle height and upwards above the chain. The inner vortices in these cases resemble those seen for rolling grain ripples\citep{mazzuoli2016on} and the oscillatory flow over a wavy wall \citep{lyne1971unsteady}. 

\begin{figure*}[ht]
    \includegraphics[width=\textwidth]{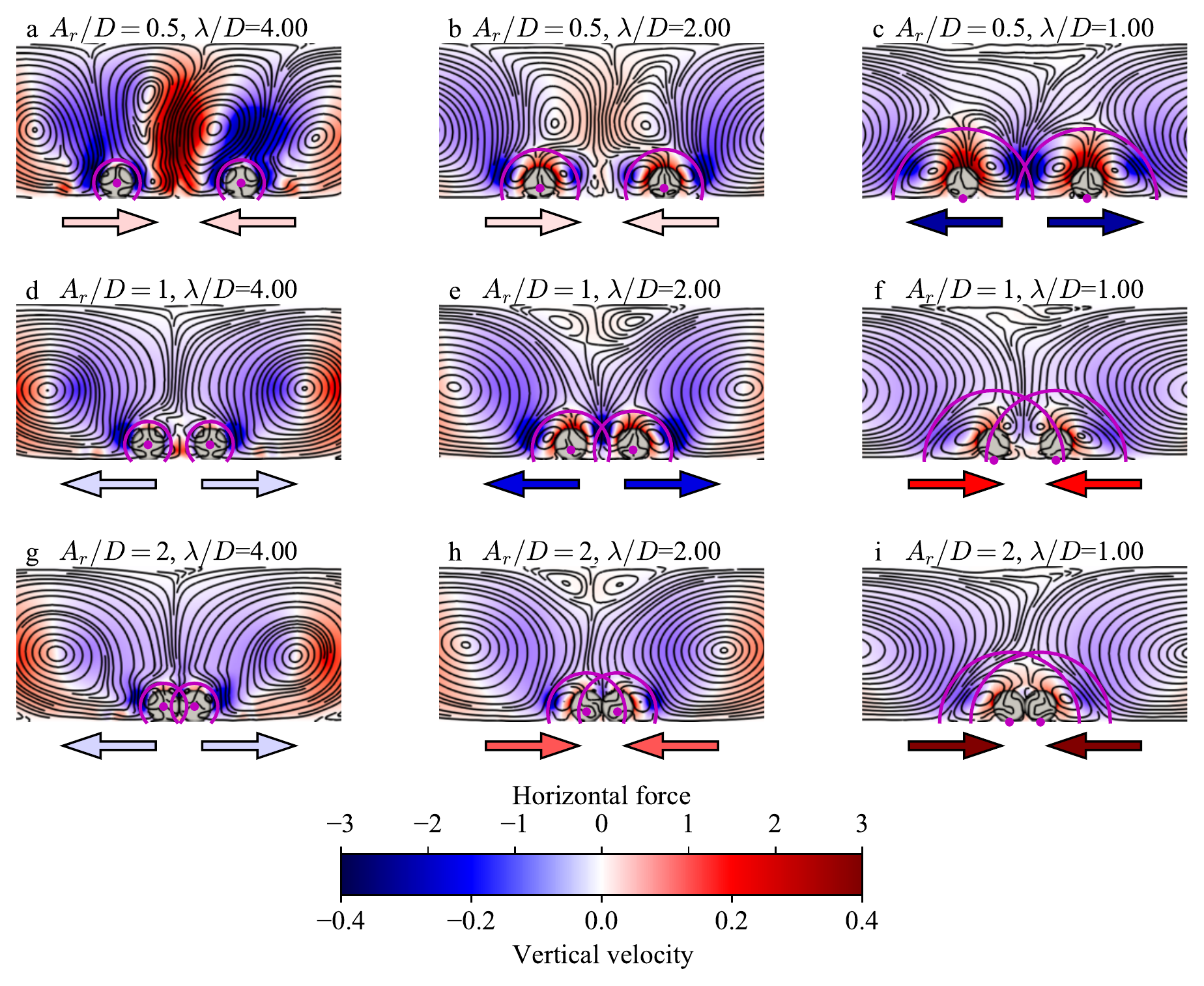}
    \caption{(Color online) The side-view of the average flow field from our numerical simulations, similar to Fig.~\ref{fig:simulationsonechain}, here for two particle chains. The distance between the chains $\lambda/D$ and the relative excursion length $A_r/D$ are both varied. The arrows indicate the direction of the (dimensionless) horizontal hydrodynamic forces on the particles, while their color is proportional to the magnitude of these forces.
    A more extensive overview with additional values of $\lambda/D$ is provided in the Supplemental Material\citep{supmat}.}
    \label{fig:simulationstwochains}
\end{figure*}

In Fig.~\ref{fig:simulationstwochains}, we present the results for simulations with two fixed chains, with different spacing between them, such that $\lambda/D=[4,2,1]$. 
In addition, we have calculated the average horizontal force on each particle, $F_x$, which is nondimensionalized by $\rho_f (\pi D^3/6) A\omega^2$ and averaged over a full oscillation period. 
An extensive overview with more values of $\lambda/D$ is given as Supplemental Material \citep{supmat}. The other parameters of the simulations are identical to those in Fig.~\ref{fig:simulationsonechain}.
The size and strength of the vortices between the two chains are highly dependent on the distance between the chains. For $A_r/D=0.5$ and relatively large spacing ($\lambda/D=4$), as shown in Fig.~\ref{fig:simulationstwochains}(a), the outer vortices induce an upward flow between the chains and a converging flow in the horizontal plane at mid-particle height. This flow results in an attractive force between the chains. 
However, as the spacing decreases ($\lambda/D=2$ and 1, in Figs.~\ref{fig:simulationstwochains}(d)~and~(g)), both the large-scale outer vortices between the chains and the upward flow disappear. This results in a (weakly) repulsive force between the chains. 

When $A_r/D=1.0$ and the chain spacing is relatively large ($\lambda/D=4$), in Fig.~\ref{fig:simulationstwochains}(b), four outer vortices and a weak upward flow between the chains are present. However, two of the outer vortices do not reach the bottom between the particles. 
As the chain spacing decreases to $\lambda/D=2$, the two outer vortices between the chains are greatly reduced, resulting in a downward flow between the chains. Correspondingly, there is a diverging flow in the horizontal mid-particle plane and a strong repulsion between the chains. However, as the spacing gets sufficiently small, i.e. as $\lambda/D=1$ in Fig.~\ref{fig:simulationstwochains}(h), the flow around the two chains becomes qualitatively similar to the flow around a single chain, as shown in Fig.~\ref{fig:simulationsonechain}(b). 
Furthermore, the force between the chains becomes strongly attractive, indicating that a two-particle-wide band is stable for these flow conditions.

For $A_r/D=2.0$ and sufficiently small spacings ($\lambda/D=1$), in Fig.~\ref{fig:simulationstwochains}(i), the flow field around the two chains is similar to that around a single chain and the force between the chains is strongly attractive. 
However, for a relatively large spacing ($\lambda/D=4$ in Fig.~\ref{fig:simulationstwochains}(c)), the inner vortices fill the entire region between the chains, such that no outer vortices are present there. The inner vortices induce a strong repulsion between the chains, similar to the case shown in Fig.~\ref{fig:simulationstwochains}(e).

In summary, the interplay of the inner and outer vortices induces attractive and repulsive hydrodynamic interactions between the chains. For a large spacing, the outer vortices generate a converging flow in the horizontal mid-particle plane, resulting in an attractive force. However, as the spacing decreases to intermediate values, these outer vortices do not fit between the chains, leading to a diverging flow and a strong repulsion. Finally, when both the spacing is small and $A_r/D$ sufficiently large, the chains strongly attract each other, with a flow resembling that around a single chain. 

The configuration where two chains touch each other ($\lambda/D=1$) is not always stable. For sufficiently low values of $A_r/D$, the chains repel each other, such as in Fig.~\ref{fig:simulationstwochains}(g). 
The figure suggests that there is a critical value of $A_r/D$ between 0.5 and 1.0, at which the net force between the chains changes from repulsive to attractive. Below this critical value, the chains always repel each other, whereas above it, the chains attract each other if they are sufficiently close. 

\begin{figure}[ht]
    \includegraphics[width=\linewidth]{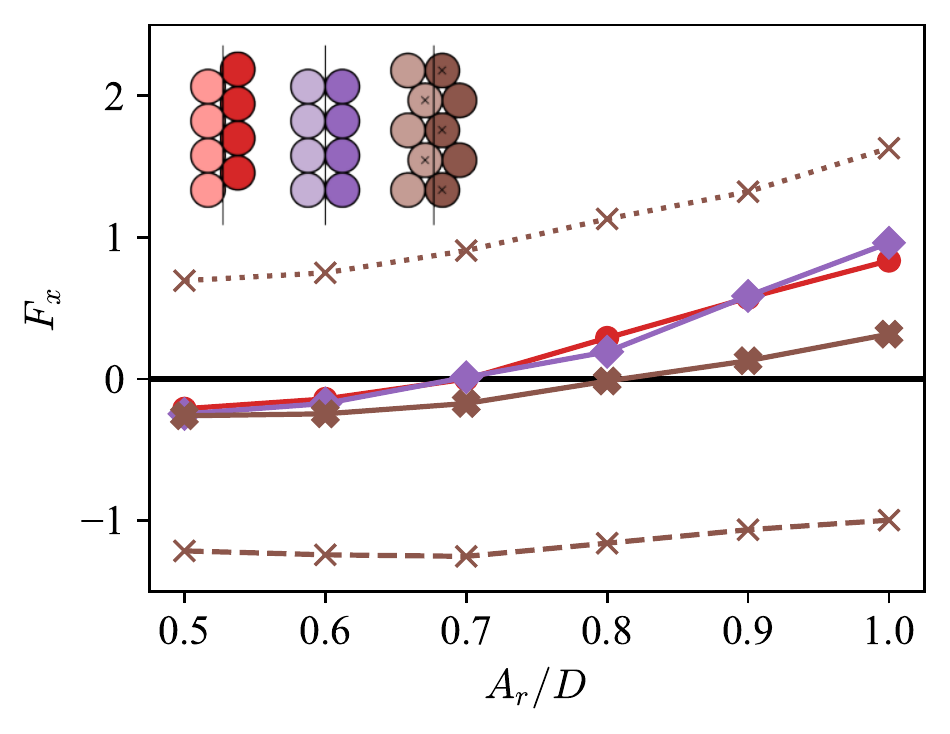}
    \caption{(Color online) The average value of the hydrodynamic force in the oscillation direction, $F_x$ (nondimensionalized by $\rho_f (\pi D^3/6) A\omega^2$), on each particle for different configurations (as indicated by the different colors) and different values of $A_r/D$. The vertical black lines show the centerlines of each configuration. Positive values correspond to (attractive) forces toward the centerline of the configuration, and thus to a stable configuration. Negative values correspond to (repulsive) forces away from the centerline of the configuration, and thus to an unstable configuration. 
    Note that the particles in the brown configuration are additionally split into inner particles (with a cross) and outer particles (unmarked), with the average forces on these particles represented by the dashed and dotted lines, respectively.}
    \label{fig:simulationforces}
\end{figure}

To get more insight into the net forces between the chains, we have performed additional numerical simulations with two rows of particles. The position of the particles is fixed, and the average streamwise force is calculated for different values of $A_r/D$ ranging from 0.5 to 1.0. The outcomes of these simulations are shown in Fig.~\ref{fig:simulationforces}. We have used three different configurations, each two rows of particles wide, based on the observations from the experiments.

The red configuration represents a two-particle-wide band with hexagonal packing. The purple configuration represents two touching chains that are aligned in the spanwise direction, similar to the alignment of the particles in two neighboring chains. The brown configuration represents a band with hexagonal packing, in the orientation that we observe in our experiment near a side wall.
For all three configurations, the streamwise force as a function of $A_r/D$ exhibits similar behavior, with repulsion at low values of $A_r/D$ and attraction for higher values. The transition from repulsion to attraction, i.e. the zero-crossing in Fig.~\ref{fig:simulationforces}, occurs at $A_r/D\approx0.7$ for the red and purple configurations, and at $A_r/D\approx0.8$ for the brown configuration.
    
For the brown configuration, we additionally distinguish between the inner and outer particles based on their distance from the centerline. We have calculated the average forces acting on these particles, where a positive value indicates a force directed towards the centerline, while a negative value indicates a force directed away from it.
The force on the outer particles (dotted line) is always directed towards the centerline. Conversely, the force on the inner particles (dashed line) is always directed away from the centerline. 
As a result, the forces on the particles in the brown configuration tend to straighten the band, leading to the purple configuration. 

Despite the expectation that the brown configuration always tends to straighten, we have frequently observed it in our experiment, particularly near the side walls, as shown in Fig.~\ref{fig:patternshapes}. 
However, for the transition from the brown to the purple configuration, the hexagonal packing should change to a square packing, requiring the particles to move in the $y$-direction (parallel to the band). Hence, there must be sufficient space for the particles to move and reorient during the reorganization. It is likely that the presence of a wall or clusters of other particles inhibits such a reorganization.

\section{\label{sec:phasespace}Overview of the parameter space}
\begin{figure*}[ht]
    \includegraphics[width=\linewidth]{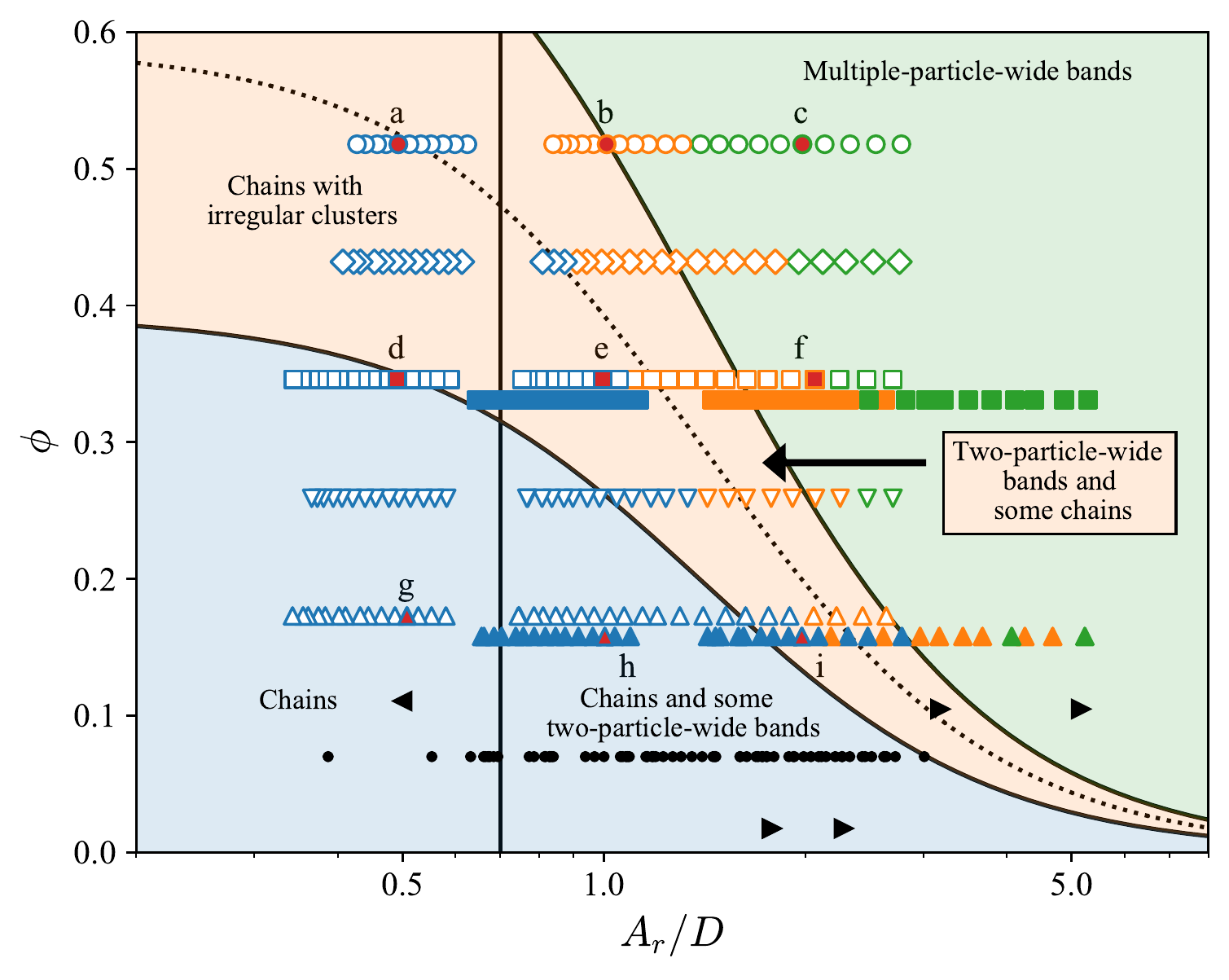}
    \caption{(Color online) 
    The location of all our experiments within the $(A_r/D,\phi)$ parameter space. The data are colored based on which fraction is largest: $F_2$ (blue), $F_4$ (orange), or $F_6$ (green). The vertical black line corresponds to $A_r/D=0.7$ and the solid black curves to $\phi[2+(A_r/D)^2]=[\pi/4,\pi/2]$ (see Eqs.~\eqref{eq:ArDcrit}~and~\eqref{eq:ArDcrit2}). The dotted black curve lies halfway between the two transitions.
    The data for the two lowest values of $\phi$ (B1 and B2, filled symbols) are slightly shifted towards lower $\phi$ values for clarity.
    The black symbols represent data from previous studies by  \citet{wunenburger2002periodic} (dots), \citet{klotsa2009chain} (left-pointing triangle), and \citet{mazzuoli2016on} (right-pointing triangles). The red symbols (labeled a-i) indicate the cases shown in Fig.~\ref{fig:patterns_phasespace}.
    }
    \label{fig:phasespace}
\end{figure*}

Based on our experiments, we have identified two transitions: from chains to two-particle-wide bands and from two-particle-wide bands to multiple-particle-wide bands. These transitions, as described by Eqs.~\eqref{eq:ArDcrit}~and~\eqref{eq:ArDcrit2}, are based on the intrinsic spacing between the chains.
Furthermore, our numerical simulations have revealed that for $A_r/D\lesssim0.7$, the two-particle-wide bands are unstable.
These conditions divide the parameter space, represented by the ($A_r/D,\phi$)-plane, into different regions, as illustrated in Fig.~\ref{fig:phasespace}. In this figure, we show the positions of all our experiments in the parameter space, with each experiment colored according to the most common number of nearest neighbors (two, four, or six), corresponding to Fig.~\ref{fig:chainwidth}.

The transition from one-particle-thick chains (blue) to two-particle-wide bands (orange) in Fig.~\ref{fig:phasespace} occurs approximately at the dotted curve, which lies approximately halfway between the solid black curves. This transition corresponds to the intersection of $F_2$ and $F_4$ halfway between the two curves, as previously shown in Fig.~\ref{fig:chainwidth}(b). Furthermore, it is worth noting that two-particle-wide bands are absent when $A_r/D<0.7$.

\begin{figure*}[ht]
    \includegraphics[width=\linewidth]{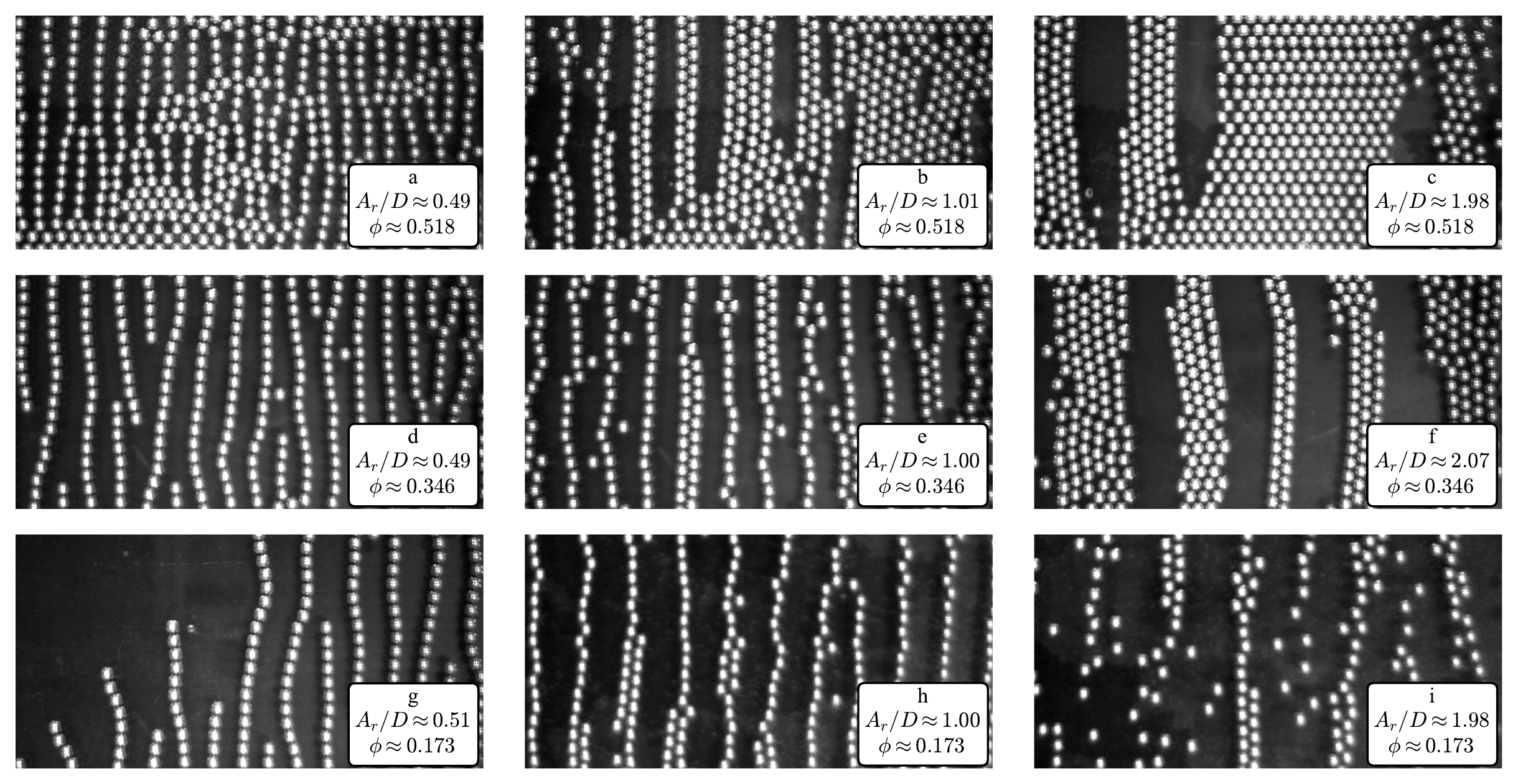}
    \caption{An overview of the patterns in our experiments for the nine cases indicated by red dots in Fig.~\ref{fig:phasespace}.}
    \label{fig:patterns_phasespace}
\end{figure*}

We now elaborate on the different regions inside the parameter space, using the experiments indicated with red dots in Fig.~\ref{fig:phasespace}. Snapshots of these experiments are shown in Fig.~\ref{fig:patterns_phasespace}. Each image is a typical example of the pattern found in the corresponding region of the parameter space. 

In the first column of Fig.~\ref{fig:patterns_phasespace} (corresponding to $A_r/D\approx0.5$), two-particle-wide bands are unstable and therefore absent. Even at high values of $\phi$, the particles do not form two-particle-wide bands but rather cluster together in irregular shapes. While some of these clusters have hexagonal packing, they lack a uniform orientation, whereas the two-particle-wide bands are oriented perpendicularly to the oscillation direction.

In the second column (corresponding to $A_r/D\approx1.0$), the two-particle-wide bands are always stable, but their presence is not always necessary to fit the pattern if $\phi$ is sufficiently small. However, depending on the initial conditions and the local particle number density, some two-particle-wide bands can form and be stable, such as shown in Fig.~\ref{fig:patterns_phasespace}(h). For higher values of $\phi$, these bands are increasingly more common as they are required to fit the pattern with its preferential spacing inside the domain. Increasingly large values of $\phi$ yield increasingly wide bands when $A_r/D\gtrsim0.7$. 

In the third column (corresponding to $A_r/D\approx2.0$), the two-particle-wide bands are always stable and always required to fit the pattern with its preferential spacing, for the values of $\phi$ considered here. The typical width of the bands increases with $\phi$. However, for low values of $\phi$, e.g. as shown in Fig.~\ref{fig:patterns_phasespace}(i), we expect to find a combination of one-particle-thick chains and two-particle-wide bands. Instead, we observe a combination of bands and isolated particles with high mobility. Consequently, the fraction of defects is high, as previously noted in Fig.~\ref{fig:chainwidth}(a) for this part of the parameter space. 

We can understand the absence of one-particle-thick chains based on the interaction between two isolated particles. When this interaction is sufficiently large, they form a pair with a small gap between them, aligned perpendicularly to the oscillation direction. Such a pair is the building block for longer chains\citep{klotsa2009chain,overveld2022numerical}. 
For $A_r/D\lesssim1$, the equilibrium distance between the particles is typically small ($\propto(\delta/D)^{1.5}$). However, as $A_r/D\gtrsim2$, the gap spacing rapidly increases ($\propto(A_r/D)^3$) due to the advection of vorticity away from the pair. This leads to weaker instantaneous interactions between the particles and destabilizes the pair configuration \citep{overveld2022numerical}.
This mechanism adds a smooth transition region between $1<A_r/D<2$ in the parameter space in Fig.~\ref{fig:phasespace}. Above this transition, the interaction between pairs of particles is too weak for one-particle-thick chains to be stable. 

It should be noted that the mechanism described above becomes irrelevant as $A_r/D$ increases further, beyond the upper solid black curve (Eq.~\eqref{eq:ArDcrit2}) in Fig.~\ref{fig:phasespace}. In that region, there are no one-particle-thick chains, as they do not fit in the domain with their preferential spacing. Instead, multiple-particle-wide bands form to fit the pattern in the domain, and these bands are stable.

Figure~\ref{fig:phasespace} further shows the location in the parameter space of different data points from previous studies. Our results agree well with the data that is available from the literature. All black symbols below the lower solid black curve correspond to one-particle-thick chains, which is consistent with the regions in the parameter space as well as our own experimental observations. 
For the two right-pointing black triangles at $\phi\approx0.1$, \citet{mazzuoli2016on} reported multiple-particle-wide bands, which again agrees with our predictions for the parameter space. 

Although the regions in the parameter space that we propose in Fig.~\ref{fig:phasespace} are supported by previous studies, we cannot make a one-to-one comparison with our experimental data due to other differences. These include differences in the particle-fluid density ratio, the relative importance of particle-bottom friction, and the smoothness of the bottom. For example, the rightmost black symbol corresponds to a simulation by \citet{mazzuoli2016on} where particles moved on top of a layer of other particles. 

\section{\label{sec:conclusions}Conclusions}
In this paper, we have described the self-organization of dense spherical particles in an oscillating flow into either one-particle-thick chains or multiple-particle-wide bands, oriented perpendicular to the direction of oscillation. Our experiments using an oscillating box show different stages of the formation, including the expansion and contraction of the pattern in the direction of oscillation. These phenomena are attributed to the repulsive and attractive interactions between chains, respectively. 

We further show that the equilibrium chain spacing is an intrinsic quantity that is set by hydrodynamic interactions, which are characterized by $A_r/D$. Conversely, the particle coverage fraction $\phi$ determines the width of the chains or bands. Using the intrinsic spacing between the chains and properties of the confinement, we accurately predict the transition from chains to bands.

Direct numerical simulations, where the flow around the particles is fully resolved, show that the interplay between the inner and outer vortices in the time-averaged flow determines the interactions between chains. The equilibrium spacing follows from the balance between long-range attractive and short-range repulsive forces, attributed to the vortices in the steady streaming flow. Our simulations further reveal the physical mechanism that stabilizes the bands when $A_r/D\gtrsim0.7$. When the chains are sufficiently close together, the inner vortices between the chains are completely suppressed, such that the average flow field around such a double band is then similar to that around a single chain. 

In conclusion, our study provides an extensive understanding of the self-organization of spherical particles into patterns in an oscillating flow. 
Using insights from both experiments and numerical simulations, we have identified the key dimensionless parameters that govern the equilibrium state of the system. Our analysis includes a comprehensive overview of the parameter space, where we identify different regions and explain the transitions between them. 
The insights gained from this study provide valuable information about the characteristics of fluid-immersed patterns and the underlying physical processes. Moreover, the findings can be used to develop novel approaches for controlling and manipulating particle-laden flows, through external forcing conditions or confinement. Additionally, the diverse range of observed patterns demonstrates the potential of hydrodynamic interactions to impose soft confinement boundaries on a collection of particles through long-range attractions.

\begin{acknowledgments}
We thank NWO for the computational resources provided on Snellius (grant EINF-2132). 
We further thank Wim-Paul Breugem (Delft University of Technology) for providing us access to the numerical code. 
Finally, we thank Janne-Mieke Meijer and Wouter Ellenbroek (both Eindhoven University of Technology) for the fruitful discussions about our experimental results and analysis methods.
\end{acknowledgments}

\section*{Data Availability Statement}
The data that support the findings of this study are (soon) openly available in 4TU.ResearchData at \url{http://doi.org/10.4121/9c7664af-4684-43cf-95cc-a3fba6951e23}

\appendix
\section{\label{sec:appendix}Role of particle-bottom friction}
We study the role of particle-bottom friction on the streamwise motion of a single particle in an oscillating box filled with viscous fluid, using the numerical code by \citet{breugem2012second}. This code is also used to uncover the physical mechanism behind the chain interactions, as described in Sec.~\ref{sec:mechanism}.
In this section, a single spherical particle is placed inside a double periodic box with lateral dimensions of $20\times10$ and a height of 5 diameters. The flow is solved on a uniform Cartesian grid with a grid spacing of $D/24$ and the particles are simulated using an immersed boundary method.
We further set $A/D=0.909$, $\delta/D\approx0.0265$, $\Gamma=0.75$, and $s=7.5$, i.e. closely matching the experimental conditions for which $A_r/D=0.5$.
The value of the Coulomb friction coefficient $\mu_c$ is varied between $0.0$ and $0.5$.

\begin{figure}[ht]
    \includegraphics[width=\linewidth]{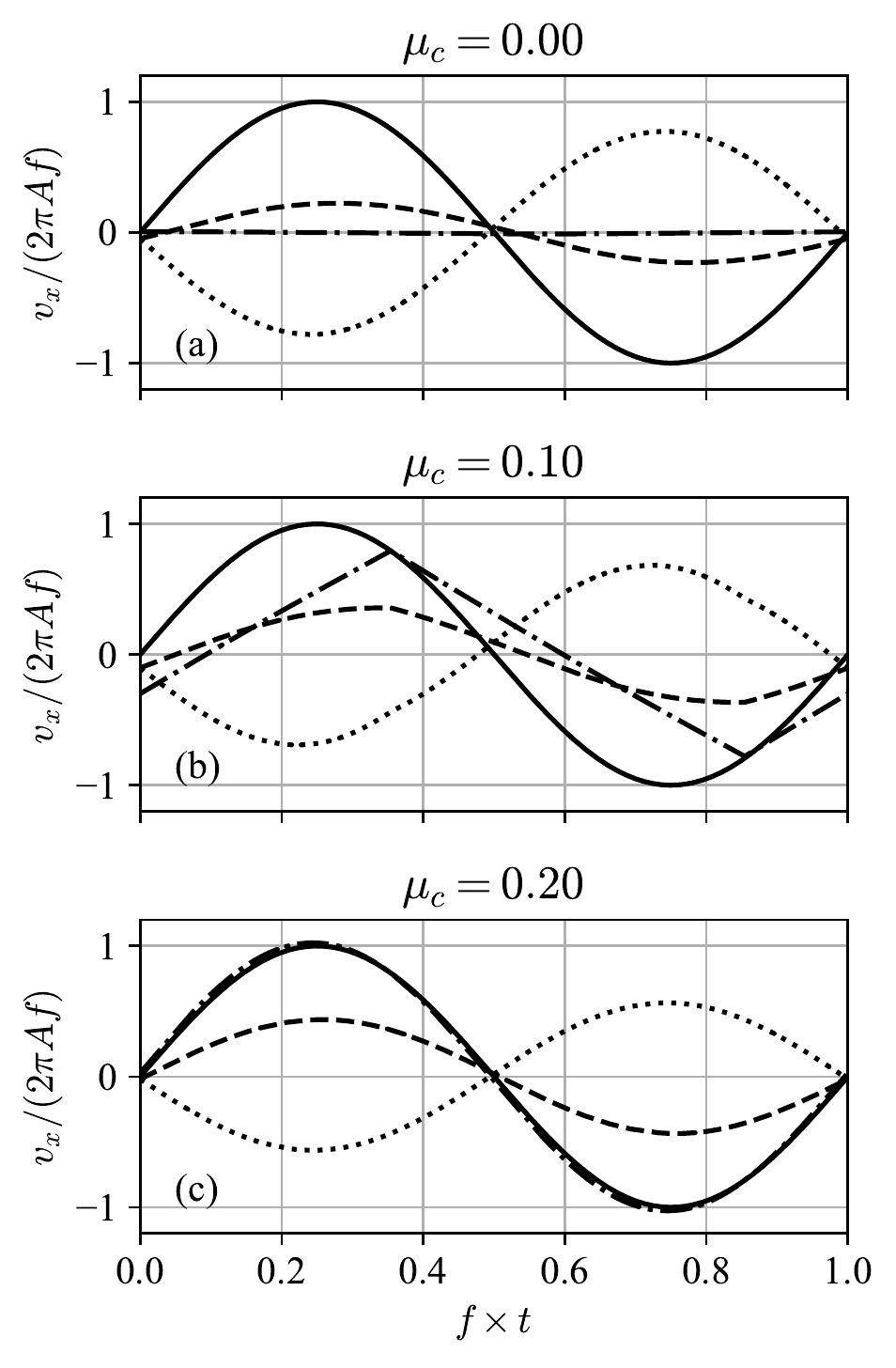}
    \caption{Velocities as a function of time, obtained from numerical simulations of a single spherical particle inside an oscillating box. The data is normalized by the velocity amplitude of the box and shown for three different values of the Coulomb friction coefficient $\mu_c$. The different lines indicate the velocity of the box (solid), the particle velocity in the lab frame (dashed), the particle velocity relative to the box (dotted), and the velocity of the point of the particle in contact with the bottom (dash-dotted). The simulations are carried out with $A/D=0.909$, $\delta/D\approx0.0265$, $\Gamma=0.75$, and $s=7.5$.}
    \label{fig:velocity_muc}
\end{figure}

Figure~\ref{fig:velocity_muc} shows, for three different values of $\mu_c$, the velocities as a function of time during the 20th oscillation period, when the system has reached a quasi-steady state and the transients have died out. The velocities are normalized by the velocity amplitude of the box, $2\pi A f$, and time is normalized by the oscillation frequency, $f$, such that one unit of time corresponds to one oscillation period.

In the absence of particle-bottom friction ($\mu_c=0.0$, in Fig.~\ref{fig:velocity_muc}(a)), the particle rotation, represented by the dash-dotted curve, is approximately zero. The hydrodynamic interactions cause slight rotation of the particle, but these effects are sufficiently small to be insignificant for the overall motion (around 0.01 on the scale used in Fig.~\ref{fig:velocity_muc}).

For slightly larger values of the Coulomb friction coefficient values, e.g. for $\mu_c=0.10$ in Fig.~\ref{fig:velocity_muc}(b), the particle-bottom friction affects both the translational and rotational motion. The particle rotates throughout the oscillatory motion, as indicated by non-zero values of the dash-dotted curve. However, the rotational velocity is not sufficiently large to match the velocity of the bottom, which is indicated by the solid curve.
When the bottom velocity exceeds that of the part of the particle in contact with the bottom, the particle experiences a positive acceleration due to the Coulomb friction. This force is proportional to $\mu_c$ but is independent of the velocity difference between the particle and bottom. Therefore, the slope (i.e. the acceleration) of the dash-dotted curve remains constant until the dash-dotted curve intersects the solid curve and the velocities match. After this point, the process reverses. This type of particle motion is commonly referred to as rolling motion with slip.     

When the Coulomb coefficient is sufficiently large, e.g. when $\mu_c=0.2$ in Fig.~\ref{fig:velocity_muc}(c), the particle rolls without slip throughout the oscillatory motion, such that the dash-dotted curve overlaps the solid curve. Note that our experiments fall into this regime. 
Increasing $\mu_c$ further does not increase the friction force or affect the particle motion. 
Finally, it is worth noting that the amplitudes of the relative particle-fluid velocities (dotted curves) decrease as friction becomes more important. 

\begin{figure}[ht]
    \includegraphics[width=\linewidth]{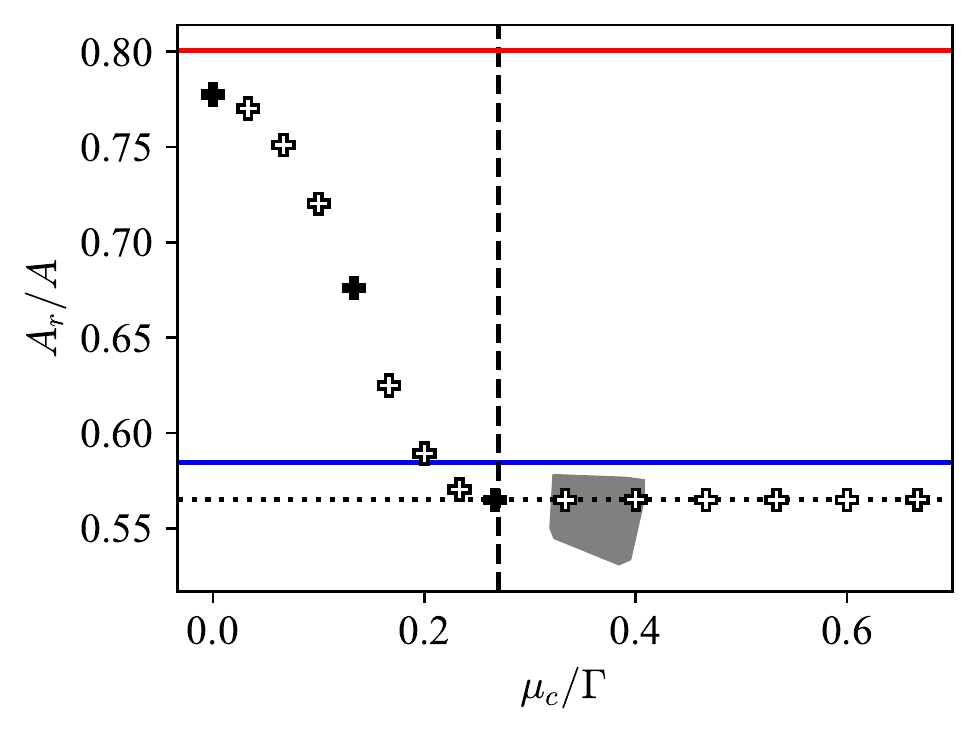}
    \caption{The ratio $A_r/A$ obtained from numerical simulations of a single spherical particle in an oscillating box. The three filled symbols correspond to the cases shown in Fig.~\ref{fig:velocity_muc}. Our experimental data fall within the gray area, where we have assumed that $\mu_c\approx0.3$.
    The dotted line at $A_r/A=0.565$ indicates the constant value of $A_r/A$ for $\mu_c/\Gamma>0.27$.
    The red and blue lines at $A_r/A\approx0.80$ and $A_r/A\approx0.58$ represent the solutions of the Basset-Boussinesq-Oseen (BBO) equation, given by Eqs.~\eqref{eq:BBO_frictionless}~and~\eqref{eq:BBO_friction}, respectively. The dashed vertical line corresponds to $\mu_c/\Gamma=0.27$ (see Eq.~\eqref{eq:rollcondition}), above which the particle rolls throughout the oscillatory motion.}
    \label{fig:ArA_mucGamma}
\end{figure}

In Fig.~\ref{fig:ArA_mucGamma}, we present our numerical simulation results for $A_r/A$ as a function of $\mu_c/\Gamma$, which sets the relative importance of particle-bottom friction\citep{overveld2022numerical}. When increasing $\mu_c/\Gamma$ from 0.0 to 0.27, there is a smooth transition from $A_r/A\approx0.778$, as the particle slides over the bottom, to $A_r/A\approx0.565$, as the particle rolls without slip. At intermediate values, the particle rolls with slip. 

For the particle to roll without slipping, the Coulomb force must always exceed the force $\boldsymbol{F}_c$ based on the inertia of the particle (see Eq.~\eqref{eq:rotationfrictionforce}). Here we have assumed that the particle rotation is solely due to particle-bottom friction. The condition for rolling without slipping is then given by
\begin{equation}\label{eq:rollcondition}
\begin{split}
    \frac{\mu_c}{\Gamma} & \geq \frac{2}{5}\frac{s}{s-1}\frac{A_r}{A}, \\
    & \gtrsim \frac{4s/5}{\sqrt{(9\delta/D)^2(2\delta/D+1)^2+(9\delta/D+14s/5+1)^2}}, \\
\end{split}
\end{equation}
where we have used the solution to the BBO equation (Eq.~\eqref{eq:BBO_friction}). For the simulations discussed in this appendix, with $\delta/D\approx0.0265$, we obtain $\mu_c/\Gamma\gtrsim0.270$, which is indicated in Fig.~\ref{fig:ArA_mucGamma}. 
The condition in Eq.~\eqref{eq:rollcondition} further confirms that the ratio $\mu_c/\Gamma$ determines the magnitude of particle-bottom friction and, consequently, the particle rotation.

Finally, we have shaded the region in Fig.~\ref{fig:ArA_mucGamma} that encompasses all our experimental data. We have estimated that $\mu_c\approx0.3$ for the interface between glass bottom and stainless steel particles \citep{engineering_toolbox}.
While the value of $\mu_c$ in the experiments may be slightly larger than 0.3, our results remain unchanged since it does not affect the relative particle-fluid motion. The value of $A_r/A$ remains constant above $\mu_c/\Gamma\approx0.27$.
Nonetheless, Fig.~\ref{fig:ArA_mucGamma} emphasizes the importance of keeping $\Gamma$ (approximately) constant in our experiments. If $\Gamma$ were varied such that $\mu_c/\Gamma\lesssim0.27$, the particle motion would transition from rolling without slipping, to rolling with slip or sliding, depending on the specific value of $\mu_c/\Gamma$.

Based on the shaded region in Fig.~\ref{fig:ArA_mucGamma}, we can conclude that in our experiments, where the relative importance of particle-bottom friction is constant, the particles always roll without slipping. As a result, keeping $\mu_c/\Gamma$ constant results in a constant value of $A_r/A$.

% \nocite{*}
% \bibliography{references}% Produces the bibliography via BibTeX.

\begin{thebibliography}{31}%
\makeatletter
\providecommand \@ifxundefined [1]{%
 \@ifx{#1\undefined}
}%
\providecommand \@ifnum [1]{%
 \ifnum #1\expandafter \@firstoftwo
 \else \expandafter \@secondoftwo
 \fi
}%
\providecommand \@ifx [1]{%
 \ifx #1\expandafter \@firstoftwo
 \else \expandafter \@secondoftwo
 \fi
}%
\providecommand \natexlab [1]{#1}%
\providecommand \enquote  [1]{``#1''}%
\providecommand \bibnamefont  [1]{#1}%
\providecommand \bibfnamefont [1]{#1}%
\providecommand \citenamefont [1]{#1}%
\providecommand \href@noop [0]{\@secondoftwo}%
\providecommand \href [0]{\begingroup \@sanitize@url \@href}%
\providecommand \@href[1]{\@@startlink{#1}\@@href}%
\providecommand \@@href[1]{\endgroup#1\@@endlink}%
\providecommand \@sanitize@url [0]{\catcode `\\12\catcode `\$12\catcode
  `\&12\catcode `\#12\catcode `\^12\catcode `\_12\catcode `\%12\relax}%
\providecommand \@@startlink[1]{}%
\providecommand \@@endlink[0]{}%
\providecommand \url  [0]{\begingroup\@sanitize@url \@url }%
\providecommand \@url [1]{\endgroup\@href {#1}{\urlprefix }}%
\providecommand \urlprefix  [0]{URL }%
\providecommand \Eprint [0]{\href }%
\providecommand \doibase [0]{http://dx.doi.org/}%
\providecommand \selectlanguage [0]{\@gobble}%
\providecommand \bibinfo  [0]{\@secondoftwo}%
\providecommand \bibfield  [0]{\@secondoftwo}%
\providecommand \translation [1]{[#1]}%
\providecommand \BibitemOpen [0]{}%
\providecommand \bibitemStop [0]{}%
\providecommand \bibitemNoStop [0]{.\EOS\space}%
\providecommand \EOS [0]{\spacefactor3000\relax}%
\providecommand \BibitemShut  [1]{\csname bibitem#1\endcsname}%
\let\auto@bib@innerbib\@empty
%</preamble>
\bibitem [{\citenamefont {Wunenburger}, \citenamefont {Carrier},\ and\
  \citenamefont {Garrabos}(2002)}]{wunenburger2002periodic}%
  \BibitemOpen
  \bibfield  {author} {\bibinfo {author} {\bibfnamefont {R.}~\bibnamefont
  {Wunenburger}}, \bibinfo {author} {\bibfnamefont {V.}~\bibnamefont
  {Carrier}}, \ and\ \bibinfo {author} {\bibfnamefont {Y.}~\bibnamefont
  {Garrabos}},\ }\bibfield  {title} {\enquote {\bibinfo {title} {Periodic order
  induced by horizontal vibrations in a two-dimensional assembly of heavy beads
  in water},}\ }\href@noop {} {\bibfield  {journal} {\bibinfo  {journal}
  {Physics of Fluids}\ }\textbf {\bibinfo {volume} {14}},\ \bibinfo {pages}
  {2350--2359} (\bibinfo {year} {2002})}\BibitemShut {NoStop}%
\bibitem [{\citenamefont {Dangles}\ \emph {et~al.}(2008)\citenamefont
  {Dangles}, \citenamefont {Steinmann}, \citenamefont {Pierre}, \citenamefont
  {Vannier},\ and\ \citenamefont {Casas}}]{dangles2008relative}%
  \BibitemOpen
  \bibfield  {author} {\bibinfo {author} {\bibfnamefont {O.}~\bibnamefont
  {Dangles}}, \bibinfo {author} {\bibfnamefont {T.}~\bibnamefont {Steinmann}},
  \bibinfo {author} {\bibfnamefont {D.}~\bibnamefont {Pierre}}, \bibinfo
  {author} {\bibfnamefont {F.}~\bibnamefont {Vannier}}, \ and\ \bibinfo
  {author} {\bibfnamefont {J.}~\bibnamefont {Casas}},\ }\bibfield  {title}
  {\enquote {\bibinfo {title} {Relative contributions of organ shape and
  receptor arrangement to the design of cricket’s cercal system},}\
  }\href@noop {} {\bibfield  {journal} {\bibinfo  {journal} {Journal of
  Comparative Physiology A}\ }\textbf {\bibinfo {volume} {194}},\ \bibinfo
  {pages} {653--663} (\bibinfo {year} {2008})}\BibitemShut {NoStop}%
\bibitem [{\citenamefont {Van~Overveld}\ \emph
  {et~al.}(2022{\natexlab{a}})\citenamefont {Van~Overveld}, \citenamefont
  {Shajahan}, \citenamefont {Breugem}, \citenamefont {Clercx},\ and\
  \citenamefont {Duran-Matute}}]{overveld2022numerical}%
  \BibitemOpen
  \bibfield  {author} {\bibinfo {author} {\bibfnamefont {T.~J. J.~M.}\
  \bibnamefont {Van~Overveld}}, \bibinfo {author} {\bibfnamefont {M.~T.}\
  \bibnamefont {Shajahan}}, \bibinfo {author} {\bibfnamefont {W.-P.}\
  \bibnamefont {Breugem}}, \bibinfo {author} {\bibfnamefont {H.~J.~H.}\
  \bibnamefont {Clercx}}, \ and\ \bibinfo {author} {\bibfnamefont
  {M.}~\bibnamefont {Duran-Matute}},\ }\bibfield  {title} {\enquote {\bibinfo
  {title} {Numerical study of a pair of spheres in an oscillating box filled
  with viscous fluid},}\ }\href@noop {} {\bibfield  {journal} {\bibinfo
  {journal} {Physical Review Fluids}\ }\textbf {\bibinfo {volume} {7}},\
  \bibinfo {pages} {014308} (\bibinfo {year} {2022}{\natexlab{a}})}\BibitemShut
  {NoStop}%
\bibitem [{\citenamefont {Van~Overveld}\ \emph
  {et~al.}(2022{\natexlab{b}})\citenamefont {Van~Overveld}, \citenamefont
  {Breugem}, \citenamefont {Clercx},\ and\ \citenamefont
  {Duran-Matute}}]{overveld2022effect}%
  \BibitemOpen
  \bibfield  {author} {\bibinfo {author} {\bibfnamefont {T.~J. J.~M.}\
  \bibnamefont {Van~Overveld}}, \bibinfo {author} {\bibfnamefont {W.-P.}\
  \bibnamefont {Breugem}}, \bibinfo {author} {\bibfnamefont {H.~J.~H.}\
  \bibnamefont {Clercx}}, \ and\ \bibinfo {author} {\bibfnamefont
  {M.}~\bibnamefont {Duran-Matute}},\ }\bibfield  {title} {\enquote {\bibinfo
  {title} {Effect of the stokes boundary layer on the dynamics of particle
  pairs in an oscillatory flow},}\ }\href@noop {} {\bibfield  {journal}
  {\bibinfo  {journal} {Physics of Fluids}\ }\textbf {\bibinfo {volume} {34}},\
  \bibinfo {pages} {113306} (\bibinfo {year} {2022}{\natexlab{b}})}\BibitemShut
  {NoStop}%
\bibitem [{eng(2004)}]{engineering_toolbox}%
  \BibitemOpen
  \href@noop {} {\enquote {\bibinfo {title} {Friction - friction coefficients
  and calculator},}\ }\bibinfo {howpublished}
  {\url{https://www.engineeringtoolbox.com/friction-coefficients-d_778.html}}
  (\bibinfo {year} {2004}),\ \bibinfo {note} {[Accessed: March
  2023]}\BibitemShut {NoStop}%
\bibitem [{\citenamefont {Allan}\ \emph {et~al.}(2019)\citenamefont {Allan},
  \citenamefont {Caswell}, \citenamefont {Keim},\ and\ \citenamefont {van~der
  Wel}}]{allan2019trackpy}%
  \BibitemOpen
  \bibfield  {author} {\bibinfo {author} {\bibfnamefont {D.~B.}\ \bibnamefont
  {Allan}}, \bibinfo {author} {\bibfnamefont {T.}~\bibnamefont {Caswell}},
  \bibinfo {author} {\bibfnamefont {N.~C.}\ \bibnamefont {Keim}}, \ and\
  \bibinfo {author} {\bibfnamefont {C.~M.}\ \bibnamefont {van~der Wel}},\
  }\href {\doibase 10.5281/zenodo.3492186} {\enquote {\bibinfo {title}
  {soft-matter/trackpy: Trackpy v0.4.2},}\ } (\bibinfo {year}
  {2019})\BibitemShut {NoStop}%
\bibitem [{\citenamefont {Crocker}\ and\ \citenamefont
  {Grier}(1996)}]{crocker1996methods}%
  \BibitemOpen
  \bibfield  {author} {\bibinfo {author} {\bibfnamefont {J.~C.}\ \bibnamefont
  {Crocker}}\ and\ \bibinfo {author} {\bibfnamefont {D.~G.}\ \bibnamefont
  {Grier}},\ }\bibfield  {title} {\enquote {\bibinfo {title} {Methods of
  digital video microscopy for colloidal studies},}\ }\href@noop {} {\bibfield
  {journal} {\bibinfo  {journal} {Journal of colloid and interface science}\
  }\textbf {\bibinfo {volume} {179}},\ \bibinfo {pages} {298--310} (\bibinfo
  {year} {1996})}\BibitemShut {NoStop}%
\bibitem [{sup()}]{supmat}%
  \BibitemOpen
  \href@noop {} {\enquote {\bibinfo {title} {See supplemental material at link
  for a video of fig.~\ref{fig:resultoverview}, and at link for a video of
  fig.~\ref{fig:resulttransient}.}}\ }\BibitemShut {NoStop}%
\bibitem [{\citenamefont {Klotsa}\ \emph {et~al.}(2009)\citenamefont {Klotsa},
  \citenamefont {Swift}, \citenamefont {Bowley},\ and\ \citenamefont
  {King}}]{klotsa2009chain}%
  \BibitemOpen
  \bibfield  {author} {\bibinfo {author} {\bibfnamefont {D.}~\bibnamefont
  {Klotsa}}, \bibinfo {author} {\bibfnamefont {M.~R.}\ \bibnamefont {Swift}},
  \bibinfo {author} {\bibfnamefont {R.~M.}\ \bibnamefont {Bowley}}, \ and\
  \bibinfo {author} {\bibfnamefont {P.~J.}\ \bibnamefont {King}},\ }\bibfield
  {title} {\enquote {\bibinfo {title} {Chain formation of spheres in
  oscillatory fluid flows},}\ }\href@noop {} {\bibfield  {journal} {\bibinfo
  {journal} {Physical Review E}\ }\textbf {\bibinfo {volume} {79}},\ \bibinfo
  {pages} {021302} (\bibinfo {year} {2009})}\BibitemShut {NoStop}%
\bibitem [{\citenamefont {Klotsa}\ \emph {et~al.}(2007)\citenamefont {Klotsa},
  \citenamefont {Swift}, \citenamefont {Bowley},\ and\ \citenamefont
  {King}}]{klotsa2007interaction}%
  \BibitemOpen
  \bibfield  {author} {\bibinfo {author} {\bibfnamefont {D.}~\bibnamefont
  {Klotsa}}, \bibinfo {author} {\bibfnamefont {M.~R.}\ \bibnamefont {Swift}},
  \bibinfo {author} {\bibfnamefont {R.~M.}\ \bibnamefont {Bowley}}, \ and\
  \bibinfo {author} {\bibfnamefont {P.~J.}\ \bibnamefont {King}},\ }\bibfield
  {title} {\enquote {\bibinfo {title} {Interaction of spheres in oscillatory
  fluid flows},}\ }\href@noop {} {\bibfield  {journal} {\bibinfo  {journal}
  {Physical Review E}\ }\textbf {\bibinfo {volume} {76}},\ \bibinfo {pages}
  {056314} (\bibinfo {year} {2007})}\BibitemShut {NoStop}%
\bibitem [{\citenamefont {Corrsin}\ and\ \citenamefont
  {Lumley}(1956)}]{corrsin1956equation}%
  \BibitemOpen
  \bibfield  {author} {\bibinfo {author} {\bibfnamefont {S.}~\bibnamefont
  {Corrsin}}\ and\ \bibinfo {author} {\bibfnamefont {J.}~\bibnamefont
  {Lumley}},\ }\bibfield  {title} {\enquote {\bibinfo {title} {On the equation
  of motion for a particle in turbulent fluid},}\ }\href@noop {} {\bibfield
  {journal} {\bibinfo  {journal} {Applied Scientific Research, Section A}\
  }\textbf {\bibinfo {volume} {6}},\ \bibinfo {pages} {114--116} (\bibinfo
  {year} {1956})}\BibitemShut {NoStop}%
\bibitem [{\citenamefont {Lotito}\ and\ \citenamefont
  {Zambelli}(2020)}]{lotito2020pattern}%
  \BibitemOpen
  \bibfield  {author} {\bibinfo {author} {\bibfnamefont {V.}~\bibnamefont
  {Lotito}}\ and\ \bibinfo {author} {\bibfnamefont {T.}~\bibnamefont
  {Zambelli}},\ }\bibfield  {title} {\enquote {\bibinfo {title} {Pattern
  detection in colloidal assembly: A mosaic of analysis techniques},}\
  }\href@noop {} {\bibfield  {journal} {\bibinfo  {journal} {Advances in
  Colloid and Interface Science}\ }\textbf {\bibinfo {volume} {284}},\ \bibinfo
  {pages} {102252} (\bibinfo {year} {2020})}\BibitemShut {NoStop}%
\bibitem [{\citenamefont {Besseling}\ \emph {et~al.}(2012)\citenamefont
  {Besseling}, \citenamefont {Hermes}, \citenamefont {Fortini}, \citenamefont
  {Dijkstra}, \citenamefont {Imhof},\ and\ \citenamefont
  {Van~Blaaderen}}]{besseling2012oscillatory}%
  \BibitemOpen
  \bibfield  {author} {\bibinfo {author} {\bibfnamefont {T.}~\bibnamefont
  {Besseling}}, \bibinfo {author} {\bibfnamefont {M.}~\bibnamefont {Hermes}},
  \bibinfo {author} {\bibfnamefont {A.}~\bibnamefont {Fortini}}, \bibinfo
  {author} {\bibfnamefont {M.}~\bibnamefont {Dijkstra}}, \bibinfo {author}
  {\bibfnamefont {A.}~\bibnamefont {Imhof}}, \ and\ \bibinfo {author}
  {\bibfnamefont {A.}~\bibnamefont {Van~Blaaderen}},\ }\bibfield  {title}
  {\enquote {\bibinfo {title} {Oscillatory shear-induced 3d crystalline order
  in colloidal hard-sphere fluids},}\ }\href@noop {} {\bibfield  {journal}
  {\bibinfo  {journal} {Soft Matter}\ }\textbf {\bibinfo {volume} {8}},\
  \bibinfo {pages} {6931--6939} (\bibinfo {year} {2012})}\BibitemShut {NoStop}%
\bibitem [{\citenamefont {Breugem}(2012)}]{breugem2012second}%
  \BibitemOpen
  \bibfield  {author} {\bibinfo {author} {\bibfnamefont {W.-P.}\ \bibnamefont
  {Breugem}},\ }\bibfield  {title} {\enquote {\bibinfo {title} {A second-order
  accurate immersed boundary method for fully resolved simulations of
  particle-laden flows},}\ }\href@noop {} {\bibfield  {journal} {\bibinfo
  {journal} {Journal of Computational Physics}\ }\textbf {\bibinfo {volume}
  {231}},\ \bibinfo {pages} {4469--4498} (\bibinfo {year} {2012})}\BibitemShut
  {NoStop}%
\bibitem [{\citenamefont {Mazzuoli}\ \emph {et~al.}(2016)\citenamefont
  {Mazzuoli}, \citenamefont {Kidanemariam}, \citenamefont {Blondeaux},
  \citenamefont {Vittori},\ and\ \citenamefont {Uhlmann}}]{mazzuoli2016on}%
  \BibitemOpen
  \bibfield  {author} {\bibinfo {author} {\bibfnamefont {M.}~\bibnamefont
  {Mazzuoli}}, \bibinfo {author} {\bibfnamefont {A.~G.}\ \bibnamefont
  {Kidanemariam}}, \bibinfo {author} {\bibfnamefont {P.}~\bibnamefont
  {Blondeaux}}, \bibinfo {author} {\bibfnamefont {G.}~\bibnamefont {Vittori}},
  \ and\ \bibinfo {author} {\bibfnamefont {M.}~\bibnamefont {Uhlmann}},\
  }\bibfield  {title} {\enquote {\bibinfo {title} {On the formation of sediment
  chains in an oscillatory boundary layer},}\ }\href@noop {} {\bibfield
  {journal} {\bibinfo  {journal} {Journal of Fluid Mechanics}\ }\textbf
  {\bibinfo {volume} {789}},\ \bibinfo {pages} {461--480} (\bibinfo {year}
  {2016})}\BibitemShut {NoStop}%
\bibitem [{\citenamefont {Lyne}(1971)}]{lyne1971unsteady}%
  \BibitemOpen
  \bibfield  {author} {\bibinfo {author} {\bibfnamefont {W.}~\bibnamefont
  {Lyne}},\ }\bibfield  {title} {\enquote {\bibinfo {title} {Unsteady viscous
  flow over a wavy wall},}\ }\href@noop {} {\bibfield  {journal} {\bibinfo
  {journal} {Journal of Fluid Mechanics}\ }\textbf {\bibinfo {volume} {50}},\
  \bibinfo {pages} {33--48} (\bibinfo {year} {1971})}\BibitemShut {NoStop}%
\bibitem [{\citenamefont {Aranson}\ and\ \citenamefont
  {Tsimring}(2006)}]{aranson2006patterns}%
  \BibitemOpen
  \bibfield  {author} {\bibinfo {author} {\bibfnamefont {I.~S.}\ \bibnamefont
  {Aranson}}\ and\ \bibinfo {author} {\bibfnamefont {L.~S.}\ \bibnamefont
  {Tsimring}},\ }\bibfield  {title} {\enquote {\bibinfo {title} {Patterns and
  collective behavior in granular media: Theoretical concepts},}\ }\href@noop
  {} {\bibfield  {journal} {\bibinfo  {journal} {Reviews of modern physics}\
  }\textbf {\bibinfo {volume} {78}},\ \bibinfo {pages} {641} (\bibinfo {year}
  {2006})}\BibitemShut {NoStop}%
\bibitem [{\citenamefont {Thomas}\ and\ \citenamefont
  {Gollub}(2004)}]{thomas2004structures}%
  \BibitemOpen
  \bibfield  {author} {\bibinfo {author} {\bibfnamefont {C.~C.}\ \bibnamefont
  {Thomas}}\ and\ \bibinfo {author} {\bibfnamefont {J.~P.}\ \bibnamefont
  {Gollub}},\ }\bibfield  {title} {\enquote {\bibinfo {title} {Structures and
  chaotic fluctuations of granular clusters in a vibrated fluid layer},}\
  }\href@noop {} {\bibfield  {journal} {\bibinfo  {journal} {Physical Review
  E}\ }\textbf {\bibinfo {volume} {70}},\ \bibinfo {pages} {061305} (\bibinfo
  {year} {2004})}\BibitemShut {NoStop}%
\bibitem [{\citenamefont {Voth}\ \emph {et~al.}(2002)\citenamefont {Voth},
  \citenamefont {Bigger}, \citenamefont {Buckley}, \citenamefont {Losert},
  \citenamefont {Brenner}, \citenamefont {Stone},\ and\ \citenamefont
  {Gollub}}]{voth2002ordered}%
  \BibitemOpen
  \bibfield  {author} {\bibinfo {author} {\bibfnamefont {G.~A.}\ \bibnamefont
  {Voth}}, \bibinfo {author} {\bibfnamefont {B.}~\bibnamefont {Bigger}},
  \bibinfo {author} {\bibfnamefont {M.~R.}\ \bibnamefont {Buckley}}, \bibinfo
  {author} {\bibfnamefont {W.}~\bibnamefont {Losert}}, \bibinfo {author}
  {\bibfnamefont {M.~P.}\ \bibnamefont {Brenner}}, \bibinfo {author}
  {\bibfnamefont {H.~A.}\ \bibnamefont {Stone}}, \ and\ \bibinfo {author}
  {\bibfnamefont {J.~P.}\ \bibnamefont {Gollub}},\ }\bibfield  {title}
  {\enquote {\bibinfo {title} {Ordered clusters and dynamical states of
  particles in a vibrated fluid},}\ }\href@noop {} {\bibfield  {journal}
  {\bibinfo  {journal} {Physical review letters}\ }\textbf {\bibinfo {volume}
  {88}},\ \bibinfo {pages} {234301} (\bibinfo {year} {2002})}\BibitemShut
  {NoStop}%
\bibitem [{\citenamefont {Jaeger}, \citenamefont {Nagel},\ and\ \citenamefont
  {Behringer}(1996)}]{jaeger1996granular}%
  \BibitemOpen
  \bibfield  {author} {\bibinfo {author} {\bibfnamefont {H.~M.}\ \bibnamefont
  {Jaeger}}, \bibinfo {author} {\bibfnamefont {S.~R.}\ \bibnamefont {Nagel}}, \
  and\ \bibinfo {author} {\bibfnamefont {R.~P.}\ \bibnamefont {Behringer}},\
  }\bibfield  {title} {\enquote {\bibinfo {title} {Granular solids, liquids,
  and gases},}\ }\href@noop {} {\bibfield  {journal} {\bibinfo  {journal}
  {Reviews of modern physics}\ }\textbf {\bibinfo {volume} {68}},\ \bibinfo
  {pages} {1259} (\bibinfo {year} {1996})}\BibitemShut {NoStop}%
\bibitem [{\citenamefont {Petit}\ and\ \citenamefont
  {Noetinger}(1988)}]{petit1988shear}%
  \BibitemOpen
  \bibfield  {author} {\bibinfo {author} {\bibfnamefont {L.}~\bibnamefont
  {Petit}}\ and\ \bibinfo {author} {\bibfnamefont {B.}~\bibnamefont
  {Noetinger}},\ }\bibfield  {title} {\enquote {\bibinfo {title} {Shear-induced
  structures in macroscopic dispersions},}\ }\href@noop {} {\bibfield
  {journal} {\bibinfo  {journal} {Rheologica acta}\ }\textbf {\bibinfo {volume}
  {27}},\ \bibinfo {pages} {437--441} (\bibinfo {year} {1988})}\BibitemShut
  {NoStop}%
\bibitem [{\citenamefont {Petit}\ and\ \citenamefont
  {Gondret}(1992)}]{petit1992redressement}%
  \BibitemOpen
  \bibfield  {author} {\bibinfo {author} {\bibfnamefont {L.}~\bibnamefont
  {Petit}}\ and\ \bibinfo {author} {\bibfnamefont {P.}~\bibnamefont
  {Gondret}},\ }\bibfield  {title} {\enquote {\bibinfo {title} {Redressement
  d'un {\'e}coulement alternatif},}\ }\href@noop {} {\bibfield  {journal}
  {\bibinfo  {journal} {Journal de Physique II}\ }\textbf {\bibinfo {volume}
  {2}},\ \bibinfo {pages} {2115--2144} (\bibinfo {year} {1992})}\BibitemShut
  {NoStop}%
\bibitem [{\citenamefont {Ouyang}\ \emph {et~al.}(2020)\citenamefont {Ouyang},
  \citenamefont {Wu}, \citenamefont {Wang}, \citenamefont {Qi}, \citenamefont
  {Li}, \citenamefont {Wang},\ and\ \citenamefont {Lu}}]{ouyang2020effect}%
  \BibitemOpen
  \bibfield  {author} {\bibinfo {author} {\bibfnamefont {L.}~\bibnamefont
  {Ouyang}}, \bibinfo {author} {\bibfnamefont {Z.}~\bibnamefont {Wu}}, \bibinfo
  {author} {\bibfnamefont {J.}~\bibnamefont {Wang}}, \bibinfo {author}
  {\bibfnamefont {X.}~\bibnamefont {Qi}}, \bibinfo {author} {\bibfnamefont
  {Q.}~\bibnamefont {Li}}, \bibinfo {author} {\bibfnamefont {J.}~\bibnamefont
  {Wang}}, \ and\ \bibinfo {author} {\bibfnamefont {S.}~\bibnamefont {Lu}},\
  }\bibfield  {title} {\enquote {\bibinfo {title} {The effect of solid content
  on the rheological properties and microstructures of a li-ion battery cathode
  slurry},}\ }\href@noop {} {\bibfield  {journal} {\bibinfo  {journal} {RSC
  advances}\ }\textbf {\bibinfo {volume} {10}},\ \bibinfo {pages}
  {19360--19370} (\bibinfo {year} {2020})}\BibitemShut {NoStop}%
\bibitem [{\citenamefont {Zhang}, \citenamefont {Snezhko},\ and\ \citenamefont
  {Sokolov}(2022)}]{zhang2022guiding}%
  \BibitemOpen
  \bibfield  {author} {\bibinfo {author} {\bibfnamefont {B.}~\bibnamefont
  {Zhang}}, \bibinfo {author} {\bibfnamefont {A.}~\bibnamefont {Snezhko}}, \
  and\ \bibinfo {author} {\bibfnamefont {A.}~\bibnamefont {Sokolov}},\
  }\bibfield  {title} {\enquote {\bibinfo {title} {Guiding self-assembly of
  active colloids by temporal modulation of activity},}\ }\href@noop {}
  {\bibfield  {journal} {\bibinfo  {journal} {Physical Review Letters}\
  }\textbf {\bibinfo {volume} {128}},\ \bibinfo {pages} {018004} (\bibinfo
  {year} {2022})}\BibitemShut {NoStop}%
\bibitem [{\citenamefont {S{\'a}nchez}, \citenamefont {Swift},\ and\
  \citenamefont {King}(2004)}]{sanchez2004stripe}%
  \BibitemOpen
  \bibfield  {author} {\bibinfo {author} {\bibfnamefont {P.}~\bibnamefont
  {S{\'a}nchez}}, \bibinfo {author} {\bibfnamefont {M.~R.}\ \bibnamefont
  {Swift}}, \ and\ \bibinfo {author} {\bibfnamefont {P.~J.}\ \bibnamefont
  {King}},\ }\bibfield  {title} {\enquote {\bibinfo {title} {Stripe formation
  in granular mixtures due to the differential influence of drag},}\
  }\href@noop {} {\bibfield  {journal} {\bibinfo  {journal} {Physical review
  letters}\ }\textbf {\bibinfo {volume} {93}},\ \bibinfo {pages} {184302}
  (\bibinfo {year} {2004})}\BibitemShut {NoStop}%
\bibitem [{\citenamefont {Xu}\ \emph {et~al.}(2016)\citenamefont {Xu},
  \citenamefont {Wang}, \citenamefont {Fang}, \citenamefont {Fu},\ and\
  \citenamefont {Yin}}]{xu2016review}%
  \BibitemOpen
  \bibfield  {author} {\bibinfo {author} {\bibfnamefont {Z.}~\bibnamefont
  {Xu}}, \bibinfo {author} {\bibfnamefont {L.}~\bibnamefont {Wang}}, \bibinfo
  {author} {\bibfnamefont {F.}~\bibnamefont {Fang}}, \bibinfo {author}
  {\bibfnamefont {Y.}~\bibnamefont {Fu}}, \ and\ \bibinfo {author}
  {\bibfnamefont {Z.}~\bibnamefont {Yin}},\ }\bibfield  {title} {\enquote
  {\bibinfo {title} {A review on colloidal self-assembly and their
  applications},}\ }\href@noop {} {\bibfield  {journal} {\bibinfo  {journal}
  {Current Nanoscience}\ }\textbf {\bibinfo {volume} {12}},\ \bibinfo {pages}
  {725--746} (\bibinfo {year} {2016})}\BibitemShut {NoStop}%
\bibitem [{\citenamefont {Blondeaux}(1990)}]{blondeaux1990sand}%
  \BibitemOpen
  \bibfield  {author} {\bibinfo {author} {\bibfnamefont {P.}~\bibnamefont
  {Blondeaux}},\ }\bibfield  {title} {\enquote {\bibinfo {title} {Sand ripples
  under sea waves part 1. ripple formation},}\ }\href@noop {} {\bibfield
  {journal} {\bibinfo  {journal} {Journal of Fluid Mechanics}\ }\textbf
  {\bibinfo {volume} {218}},\ \bibinfo {pages} {1--17} (\bibinfo {year}
  {1990})}\BibitemShut {NoStop}%
\bibitem [{\citenamefont {Zablotsky}, \citenamefont {Blums},\ and\
  \citenamefont {Herrmann}(2017)}]{zablotsky2017self}%
  \BibitemOpen
  \bibfield  {author} {\bibinfo {author} {\bibfnamefont {D.}~\bibnamefont
  {Zablotsky}}, \bibinfo {author} {\bibfnamefont {E.}~\bibnamefont {Blums}}, \
  and\ \bibinfo {author} {\bibfnamefont {H.~J.}\ \bibnamefont {Herrmann}},\
  }\bibfield  {title} {\enquote {\bibinfo {title} {Self-assembly and rheology
  of dipolar colloids in simple shear studied using multi-particle collision
  dynamics},}\ }\href@noop {} {\bibfield  {journal} {\bibinfo  {journal} {Soft
  Matter}\ }\textbf {\bibinfo {volume} {13}},\ \bibinfo {pages} {6474--6489}
  (\bibinfo {year} {2017})}\BibitemShut {NoStop}%
\bibitem [{\citenamefont {Riley}(1966)}]{riley1966sphere}%
  \BibitemOpen
  \bibfield  {author} {\bibinfo {author} {\bibfnamefont {N.}~\bibnamefont
  {Riley}},\ }\bibfield  {title} {\enquote {\bibinfo {title} {On a sphere
  oscillating in a viscous fluid},}\ }\href@noop {} {\bibfield  {journal}
  {\bibinfo  {journal} {The Quarterly Journal of Mechanics and Applied
  Mathematics}\ }\textbf {\bibinfo {volume} {19}},\ \bibinfo {pages} {461--472}
  (\bibinfo {year} {1966})}\BibitemShut {NoStop}%
\bibitem [{\citenamefont {Fortes}, \citenamefont {Joseph},\ and\ \citenamefont
  {Lundgren}(1987)}]{fortes1987nonlinear}%
  \BibitemOpen
  \bibfield  {author} {\bibinfo {author} {\bibfnamefont {A.~F.}\ \bibnamefont
  {Fortes}}, \bibinfo {author} {\bibfnamefont {D.~D.}\ \bibnamefont {Joseph}},
  \ and\ \bibinfo {author} {\bibfnamefont {T.~S.}\ \bibnamefont {Lundgren}},\
  }\bibfield  {title} {\enquote {\bibinfo {title} {Nonlinear mechanics of
  fluidization of beds of spherical particles},}\ }\href@noop {} {\bibfield
  {journal} {\bibinfo  {journal} {Journal of Fluid Mechanics}\ }\textbf
  {\bibinfo {volume} {177}},\ \bibinfo {pages} {467--483} (\bibinfo {year}
  {1987})}\BibitemShut {NoStop}%
\bibitem [{\citenamefont {Ara{\'u}jo}\ \emph {et~al.}(2023)\citenamefont
  {Ara{\'u}jo}, \citenamefont {Janssen}, \citenamefont {Barois}, \citenamefont
  {Boffetta}, \citenamefont {Cohen}, \citenamefont {Corbetta}, \citenamefont
  {Dauchot}, \citenamefont {Dijkstra}, \citenamefont {Durham}, \citenamefont
  {Dussutour} \emph {et~al.}}]{araujo2023steering}%
  \BibitemOpen
  \bibfield  {author} {\bibinfo {author} {\bibfnamefont {N.~A.}\ \bibnamefont
  {Ara{\'u}jo}}, \bibinfo {author} {\bibfnamefont {L.}~\bibnamefont {Janssen}},
  \bibinfo {author} {\bibfnamefont {T.}~\bibnamefont {Barois}}, \bibinfo
  {author} {\bibfnamefont {G.}~\bibnamefont {Boffetta}}, \bibinfo {author}
  {\bibfnamefont {I.}~\bibnamefont {Cohen}}, \bibinfo {author} {\bibfnamefont
  {A.}~\bibnamefont {Corbetta}}, \bibinfo {author} {\bibfnamefont
  {O.}~\bibnamefont {Dauchot}}, \bibinfo {author} {\bibfnamefont
  {M.}~\bibnamefont {Dijkstra}}, \bibinfo {author} {\bibfnamefont
  {W.}~\bibnamefont {Durham}}, \bibinfo {author} {\bibfnamefont
  {A.}~\bibnamefont {Dussutour}},  \emph {et~al.},\ }\bibfield  {title}
  {\enquote {\bibinfo {title} {Steering self-organisation through
  confinement},}\ }\href@noop {} {\bibfield  {journal} {\bibinfo  {journal}
  {Soft Matter}\ } (\bibinfo {year} {2023})}\BibitemShut {NoStop}%
\end{thebibliography}

%merlin.mbs aipnum5-1.bst 2010-07-25 4.21a (PWD, AO, DPC) hacked

\end{document}